\documentclass[10pt,twocolumn,showpacs,preprintnumbers,amsmath,amssymb,prl]{revtex4-1}
\usepackage{graphicx}% Include figure files
\usepackage{epstopdf}
\usepackage{dcolumn}% Align table columns on decimal point
\usepackage{bm}% bold math
\usepackage{color}
\usepackage{forloop}
\usepackage{amsfonts}

\definecolor{blue}{RGB}{0, 0, 150}
\definecolor{green}{RGB}{0,150,0}
\definecolor{red}{RGB}{200, 0, 0}
\definecolor{black}{RGB}{0, 0, 0}
\renewcommand{\l}{\left}
\renewcommand{\r}{\right}
\newcommand{\subeq}[2]{\begin{subequations}\label{#2}\begin{align}#1\end{align}\end{subequations}}
\newcommand{\al}[2]{\begin{align}#1\label{#2}\end{align}}
\newcommand{\eq}[2]{\begin{equation}#1\label{#2}\end{equation}}

\newcommand{\figureroot}{./}

\begin{document}
\title{Bursting noise in gene expression dynamics: \\ Linking microscopic and mesoscopic models}

\author{Yen Ting Lin}
\email{yenting.lin@manchester.ac.uk}
\affiliation{Theoretical Physics, School of Physics and Astronomy, The University of 
Manchester, Manchester M13 9PL, UK}

\author{Tobias Galla}
\email{tobias.galla@manchester.ac.uk}
\affiliation{Theoretical Physics, School of Physics and Astronomy, The University of 
Manchester, Manchester M13 9PL, UK}

\date{\today}

\begin{abstract}
The dynamics of short-lived mRNA results in bursts of protein production in gene regulatory networks. We investigate the propagation of bursting noise between different levels of mathematical modelling, and demonstrate that conventional approaches based on diffusion approximations can fail to capture bursting noise. An alternative coarse-grained model, the so-called piecewise deterministic Markov process (PDMP), is seen to outperform the diffusion approximation in biologically relevant parameter regimes. We provide a systematic embedding of the PDMP model into the landscape of existing approaches, and we present analytical methods to calculate its stationary distribution and switching frequencies.
\end{abstract}
\maketitle

\newcommand{\NPI}{N_{\rm X}}
\newcommand{\NPII}{N_{\rm Y}}
\newcommand{\NPIII}{N_{\rm Z}}
\newcommand{\HMF}{H_{\rm MF}}
\newcommand{\HSIBB}{H_{}}
\newcommand{\HSIB}{B\times H_{}}
\newcommand{\HIB}{H_{}}
\newcommand{\HP}{H_\text{}}
\newcommand{\po}{p_\text{0}}
\newcommand{\pX}{p_\text{X}}
\newcommand{\pY}{p_\text{Y}}
\newcommand{\GIB}{G_\text{}}
\setcitestyle{super}

Transcription and translation in the process of gene expression occur at the molecular level and in environments of relatively small copy numbers. The discreteness of the molecular dynamics and the inherent randomness with which reactions occur are known as `intrinsic noise'. It is now widely accepted that intrinsic noise plays an important role in gene regulatory networks\cite{Kaern,Thattai,Walczak}. It promotes epigenetic diversity and enhances the adaptability of a single phenotype in changing environments \cite{Thattai2,Acar}. 
To investigate the effects of intrinsic noise, mathematical models at different levels have been constructed, ranging from microscopic models \cite{Roberts,Arkin,Walczak,WalczakSasai,Roma,Warren} describing the finer origins of intrinsic noise to mesoscopic models\cite{Wang,WangHuang,Lu,Friedman,Bokes}.
While the former capture the biological processes in more detail, the latter are computationally scalable and constructed to model more complex networks. These models all capture some signatures of intrinsic noise, but the detailed implementation of stochasticity varies from model to model. It is then important to consider how noise propagates between different levels of mathematical modelling.  At present coarse-grained models are often proposed \emph{ad hoc} and not derived from the more detailed lower-scale models.  Is this always mathematically appropriate?  What statistics of noise should modellers use at the different levels of coarse graining?  What are the consequences of the choice of noise statistics, and what are the pitfalls in deriving models on the meso-level from finer models on smaller scales?  These are some of the questions we aim to address in this work.

The above difficulties in transitioning between different levels of modelling can nicely be illustrated in the context of biological switches. These are systems with different metastable states and the possibility to `switch' between those states.  Biological organisms with such behaviour include the \emph{Lac} switch \cite{Roberts} in \emph{Escherichia coli} and the \emph{Enterobacteria phage $\lambda$} switch \cite{Arkin}. 
Computational and mathematical models of these range from very detailed descriptions \cite{Roberts, Arkin} over individual-molecule approaches \cite{Roma,WalczakSasai,Strasser,Warren} to mesoscopic models \cite{Wang,WangHuang,Friedman,Bokes}.  

\begin{figure*}
\begin{center}
\includegraphics[width=0.9\textwidth]{\figureroot 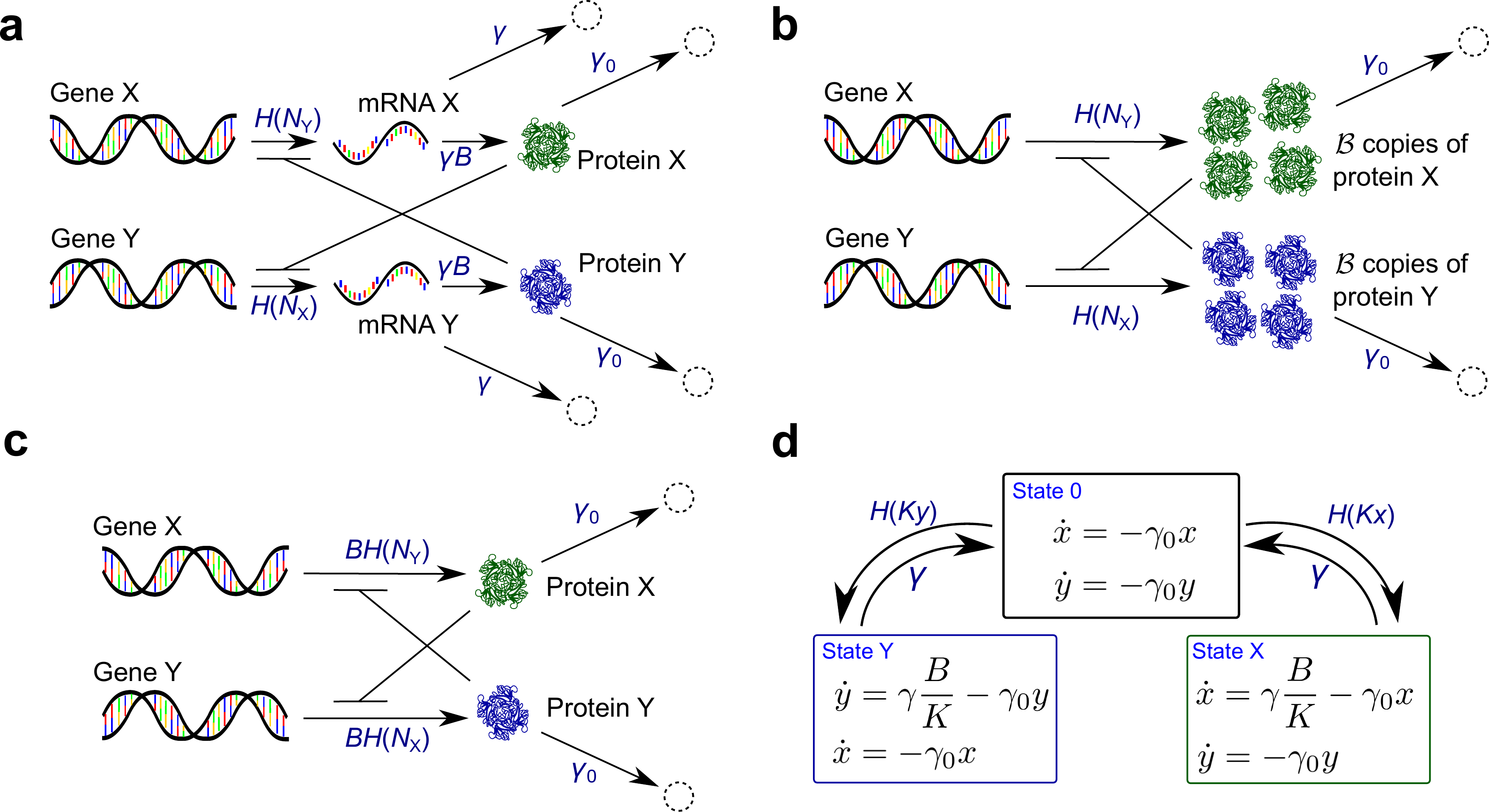}
\caption{Schematic diagrams illustrating the model dynamics. ({\bf a}) Full model (FM) describing both the mRNA and the protein populations; ({\bf b}) Protein-only model with geometrically distributed (GB) or constant (CB) bursts. The quantity $\mathcal{B}$ is a geometrically distributed random number with mean $B$ in the GB model, and $\mathcal{B}=B$ is a constant in the CB model; ({\bf c}) Protein-only model without bursts (NB); ({\bf d}) The piecewise deterministic Markov process (PDMP).} \label{fig:1}
\end{center}
\end{figure*}

The difficulties in connecting these different levels of modelling biological switches are amplified by the recent recognition that the mRNA populations are essential to describing the statistics of regulatory processes \cite{Assaf,Strasser}. 
Biologically, mRNA molecules are a relatively short-lived source compared to the proteins into which they ultimately translate. 
Protein production from a given mRNA molecule proceeds while it exists, but ceases after the mRNA decays. 
This leads to a production of protein in bursts---that is, the production is active for a relatively short and random period of mRNA lifetime, and during that time a \emph{random number} of proteins is generated. 
This phenomenon is termed \emph{translational bursting} \cite{Kaern} and it can be observed in single-molecule experiments \cite{Cai}.  
While some mesoscopic models account for such bursting \cite{Friedman,Bokes}, the theoretical investigation of these processes is often limited to their stationary distribution and frequently does not include dynamic features such as switching times.

The aim of our work is to investigate the effects bursting noise in gene regulatory networks \cite{Gardner,Roma,Strasser,JunweiWang,Warren,Wang,Lu,Lu2}, and to construct connections between individual-based models and mesoscopic approaches.  Specifically, we start from microscopic and individual-molecule-based models of a toggle switch and set out to construct coarse grained, mathematically tractable models without systematically biasing the outcomes. 

\section{Results}
\noindent{\bf Different scales of individual-based models of a toggle switch network.}
We compare four individual-based models and investigate the effect of bursting noise in a toggle switch network. 
The first model we consider describes both the mRNA and the protein population dynamics\cite{Strasser}.  
Fig.~\ref{fig:1}a illustrates the Markovian model of the regulatory network. 
Genes X and Y are transcribed into mRNA X and mRNA Y, respectively, which in turn are translated to produce proteins X and Y. 
The transcription of each of the two genes is suppressed by proteins of the respective other type via a Hill function\cite{Thattai,Walczak} $H(N)=K\left[r_0+r/[1+(N/K)^n]\right]$, where $N$ stands for the number of suppressing proteins. The model parameter $K$ represents a typical population scale of the proteins, and the parameters $r$ and $r_0$ set the minimal ($r_0K$) and maximal transcription rates ($(r_0+r)K$). 
The parameter $n>0$ is the so-called Hill coefficient which models the cooperative binding of the repressors \cite{Walczak}. 
More details of the reaction scheme can be found in the Supplementary Information. 
Proteins of either type, and the mRNA molecules degrade with constant rates $\gamma_0$ and $\gamma$ respectively.  
Biologically, mRNA molecules degrade much faster than the proteins do ($\gamma \gg \gamma_0$) \cite{Thattai,Friedman,Cai}. 
The translation rate of the mRNA is parametrised by $\gamma B$ where the parameter $B$ is the relative frequency of protein production to mRNA degradation. In this parametrisation, the number of proteins one single mRNA molecule produces during its lifetime is a geometrically distribution random variable with mean $B$ (see Supplementary Information). Biologically the parameter $B$ varies depending on the type of product protein \cite{Swain}.
We assume $B\gtrsim 10$ in this work \cite{Thattai,WalczakSasai} to investigate the effect of translational bursting.
Together with the relatively short lifetime of mRNA molecules, this constitutes the origin of `translational bursting' in the model \cite{Friedman,Raj}: a relatively large number of protein molecules is synthesized in a relatively short period of time. 

\begin{table*}[t]
\begin{center}
\begin{tabular}{|c|c|c|c|c|}
\hline
Parameter & Description & Value & Unit & Reference \\ \hline
\hline
$B$ & Average number of protein each mRNA produces  & 30 & molecule & \cite{Thattai}\\ 
$\gamma$ & mRNA degradation rate & 30 & 1/(cell cycle) & \cite{Thattai}\\
$\gamma_0$ & Protein degradation rate & 1.0 & 1/(cell cycle) & \cite{Thattai,Roma,Taniguchi}\\ 
$r$ &  Maximum suppressed transcription rate  & $ 6/100 $  &  1/(cell cycle) & \cite{Lu,Kobayashi}\footnote{In \cite{Lu} $r=1.8$ and the time unit is defined as the inverse of the protein degradation rate. In our full model we use this value, normalized by to the mean burst size $B=30$ molecules ($r=1.8/30=0.06$).} \\
$r_0$ &  Basal transcription rate  &  $1/150$ &  1/(cell cycle) & \cite{Lu,Kobayashi}\footnote{In \cite{Lu} $r_0=0.2$. After normalising with respect to the burst size $30$, we obtain $1/150$. In \cite{Kobayashi} $r_0=.05r$, which is of the same order as \cite{Lu}.} \\
$K$ & A typical population scale of the proteins & 200 &  molecule & \cite{Lu,Kobayashi}\footnote{In \cite{Lu} $K$ is set to be 200 molecules. In \cite{Kobayashi} only the deterministic dynamics are provided and $r+r_0=4.0$. To match the protein population scale $\approx 400$ in \cite{Lu,Taniguchi}, we impose $rK=400$, resulting in a typical population scale of the proteins $K~\sim 100$ molecules, which is of the same order as that of \cite{Lu}.} \\
$n$ & Hill coefficient & 3.0 & Dimensionless & \cite{Roma,Gardner,Lu,Kobayashi}\\
\hline
\end{tabular}
\caption{Parameter set.}\label{table:1}
\end{center}
\end{table*}

For simplicity, the process in Fig.~\ref{fig:1}a is assumed to be symmetric with respect to X and Y, but the analysis is easily generalised to asymmetric circuits. 
In Table \ref{table:1} we list a set of estimated values of the parameters for the model organism \emph{E. coli}, along with relevant references.

In the context of this work the model just described constitutes the most detailed model we will investigate and compare against. 
It serves as a starting point for the derivation of more coarse grained models, and for these purposes we will refer to it as the `full model' ({\bf FM}) in the following.

The FM describes both the mRNA and the protein populations, hence it constitutes a relatively high-dimensional system which complicates the mathematical analysis. Notably, the only role of mRNA in the FM is to generate proteins, and so mRNA can be left out, so long as the correct statistics of protein production is retained.  The timescale separation between the mRNA and protein lifetimes leads to the following reduced model describing only the protein dynamics.  In the limit of infinitely-fast mRNA degradation ($\gamma\gg\gamma_0$), proteins are generated instantaneously in bursts of geometrically distributed sizes with a mean $B$, and in between bursting events protein populations decay with rate $\gamma_0$. 
We will refer to the reduced model as the {\bf GB} model (geometrically distributed bursts), see Fig.~\ref{fig:1}b \cite{Swain,Assaf}. In the GB model, the transcription rates are regulated via the Hill function exactly as before in the FM.

A further reduction of the GB model involves replacing the geometrically distributed burst sizes by a constant size $B$. We will call this the {\bf CB} model (constant bursts) \cite{WalczakSasai}. While the reduction of the full model to the model with geometrically distributed bursts is well controlled and exact in the limit $\gamma\gg\gamma_0$, the effects of introducing constant burst sizes are unclear at this stage, and require a detailed analysis (see below).

An even more reduced model is a model with no bursts\cite{Warren,Roma,WalczakSasai}, we will refer to this as the {\bf NB} model. The reaction scheme is illustrated in Fig.~\ref{fig:1}c.  In this model, only one single protein is synthesized when a transcription event occurs.  We assume a $B$-fold increased transcription rate so that the average number of proteins synthesized per unit time is consistent with the FM, GB, and CB models.
\\

\begin{figure}
\begin{center}
\includegraphics[width=0.48\textwidth]{\figureroot 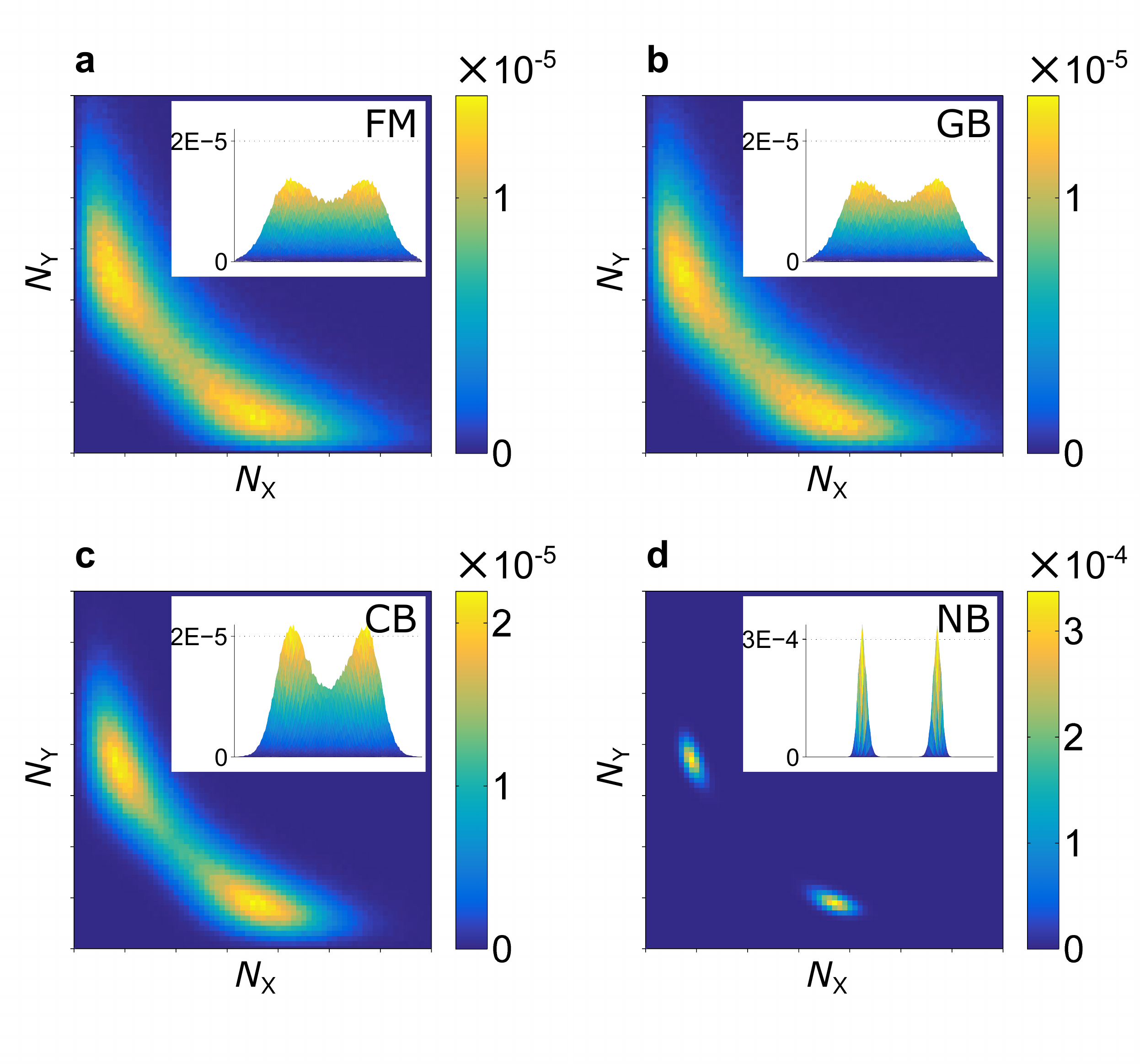}
\caption{Stationary distribution of protein numbers, shown in the range $0\leq \NPI,\NPII\leq 700$ on a linear scale on both axes. ({\bf a}) FM: Full model describing the mRNA and protein populations; ({\bf b}) GB: protein-only model with geometrically distributed bursts; ({\bf c}) CB: protein-only model with constant bursts; and ({\bf d}) NB: protein-only model without bursts. 
\label{fig:2}}
\end{center}
\end{figure}

\noindent{\bf Only the GB model approximates the stationary distribution of the FM.}
Numerical simulations of each of the models are carried out using standard methods \cite{Schwartz,Gillespie}. 
In the following we present statistical properties of the models, leaving typical sample paths to the Supplementary Information.
Fig.~\ref{fig:2} displays the numerically computed stationary distributions for the FM, GB, CB and NB models.
They illustrate that  the profiles of protein expressions in different model settings are quite distinct. This is due to the different representation of the underlying intrinsic noise.  

While the stationary distributions of the FM and the GB model are in good agreement with each other, substantial discrepancies from the full model are found in the CB and NB models. In the CB model the stationary distribution of protein numbers is very localised compared to the FM and the GB model. In the NB model the probability distribution is even more sharply concentrated. This is because the NB model misses out two pertinent sources of noise. Bursting production in the CB model amplifies the stochasticity of transcription events and leads to a broadening of the protein distribution. Adding randomly distributed burst sizes (GB model) introduces further stochasticity, and diversifies of protein numbers even further.

Based on these results, we conclude that the bursting noise introduced by the mRNA populations significantly broadens the stationary distribution.  In addition, the GB model approximates the FM model significantly better than the CB and NB models do. We can effectively discard the CB and NB models as faithful representation of the FM, and our subsequent discussion hence focuses mostly on the GB model.
\\

\noindent{\bf The GB model approximates the mean first switching time of the FM.}
The toggle switch has two dynamic attractors, one in which protein X is highly expressed and where protein Y has a low concentration, and the other with inverted roles by symmetry.  Starting from one attractor the switch can be driven to the other attractor by fluctuations. The timescale of such a transition quantifies the dynamical stability: the longer the timescale, the more stable the system is at the initial position. As we will study next, the way in which the bursting production of protein is implemented significantly affects the timescale of these switching processes. 

Starting from initial condition $\NPI(0)=n_{x,0}$ and $\NPII(0)=n_{y,0}$, we define the first switching time as the time it takes a sample path to reach the symmetric boundary $\NPI=\NPII$.  Mathematically, the first switching time is a random variable. The mean first switching time (MFST) is then the average value of the random first switching time. The MFST depends on the initial condition $\l(n_{x,0},n_{y,0}\r)$. 

Sweeping across the space of possible initial configuration, the MFST of the FM and of the GB model are measured in simulations and presented in Fig.~\ref{fig:3}.  We show the MFST of the CB in the Supplementary Information.  As with the stationary distributions, the data in Fig.~\ref{fig:3} indicates that the GB model approximates the switching times of the full model to a good accuracy.  We remark that the MFST of the CB model is almost as twice as long as that of the GB and FM models, and the switching time in the NB model is longer than $1000$ cell cycles (Supplementary Information).
\\

\noindent{\bf Diffusion approximation of the GB model.}
The evolution of the protein population in the GB model is described by a master equation (Supplementary Information). 
Solving master equations mathematically is however difficult and mostly limited to linear dynamics \cite{Kumar,Swain}. The only realistic way forward for a theoretical analysis is often the so-called diffusion approximation.

In the diffusion approximation, the discrete-molecule process is approximated by a Gaussian process for continuous concentrations---numbers of the different types of molecules normalized by a typical population scale. The Gaussian process satisfies a diffusion equation (the Fokker--Planck equation) \cite{vanKampen,Gardiner}. Based on these methods, it is often possible to calculate or approximate the stationary behaviour and switching times of model gene networks. 
For existing studies in the context of toggle switches see \cite{Wang,Lu,WangHuang}.

\begin{figure}
\begin{center}
\includegraphics[width=0.48\textwidth]{\figureroot 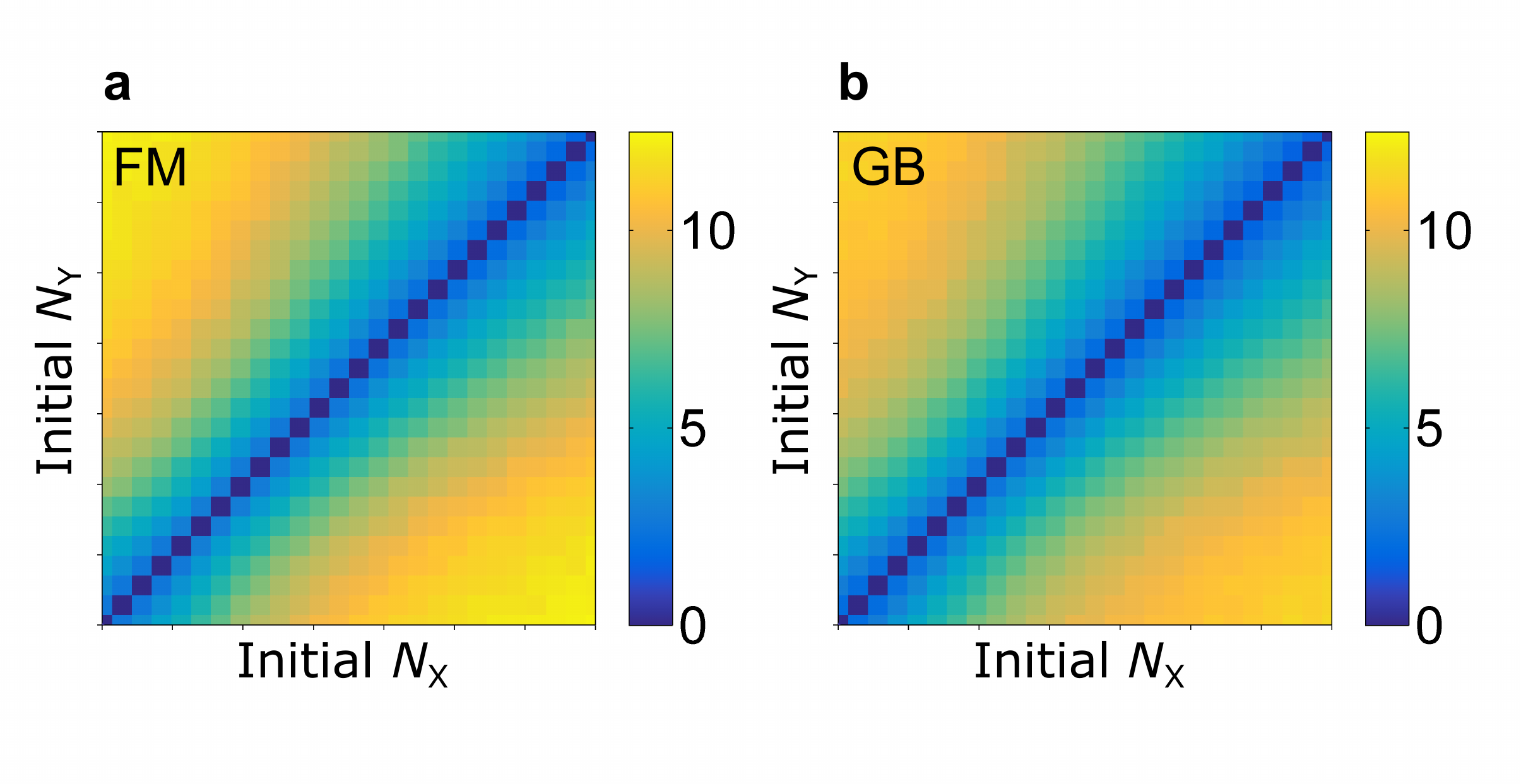}
\caption{Mean first switching time as a function of the initial protein numbers ($0\leq \NPI,\NPII\leq 700$, shown on a linear scale). ({\bf a}) FM: Full model; ({\bf b}) GB model. \label{fig:3}}
\end{center}
\end{figure}

Deriving the diffusion approximation of the GB model requires modest modifications to the standard Kramers--Moyal expansion \cite{vanKampen,Gardiner}. These modifications are necessary to account for the randomness induced by the geometrically distributed burst size. Details of the derivation can be found in the Supplementary Information, we here only report the final outcome. The expansion results in two coupled It\=o stochastic differential equations for the concentrations $x_t=\NPI(t)/K$ and $y_t=\NPII(t)/K$. These are valid in the limit of large but finite populations \cite{KurtzFP} and of the form
\subeq{
dx_t ={}& v(x_t,y_t) dt + \sqrt{D(x_t,y_t)} dW_t^{(x)}, \\
dy_t ={}& v(y_t,x_t) dt + \sqrt{D(y_t,x_t)} dW_t^{(y)}, 
}{eq:diffapprox}
with drift $v$ and diffusion $D$ given by 
\subeq{
v(w,z):={}&B \l(r_0+ \frac{r}{1+z^n}\r)-  {\gamma_0} w, \\
D(w,z):={}&  \frac{B}{K}\l[ \l(2B+1\r) \l(r_0+ \frac{r}{1+z^n}\r) +  \frac{\gamma_0}{B} w\r].
}{eq:vD}
The quantities $dW_t^{(x)}$ and $dW_t^{(y)}$ represent independent Wiener processes. 

The diffusion approximation can only be expected to be accurate when molecule numbers are large so that the concentations $x_t$ and $y_t$ are effectively continuous. In principle, a similar analysis can also be applied to the master equations of the full model. In the FM mRNA numbers are rather small though (typically $<5$, see Supplementary Information), so the Gaussian approximation does not capture the statistics of the intrinsic noise faithfully. Similarly further analysis of the CB and NB models can be carried out based on the diffusion approximation.  Given that CB and NB models fail to reproduce the behaviour of the FM, these results are relegated to the Supplementary Information.
 
\begin{figure}
\begin{center}
\includegraphics[width=0.48\textwidth]{\figureroot 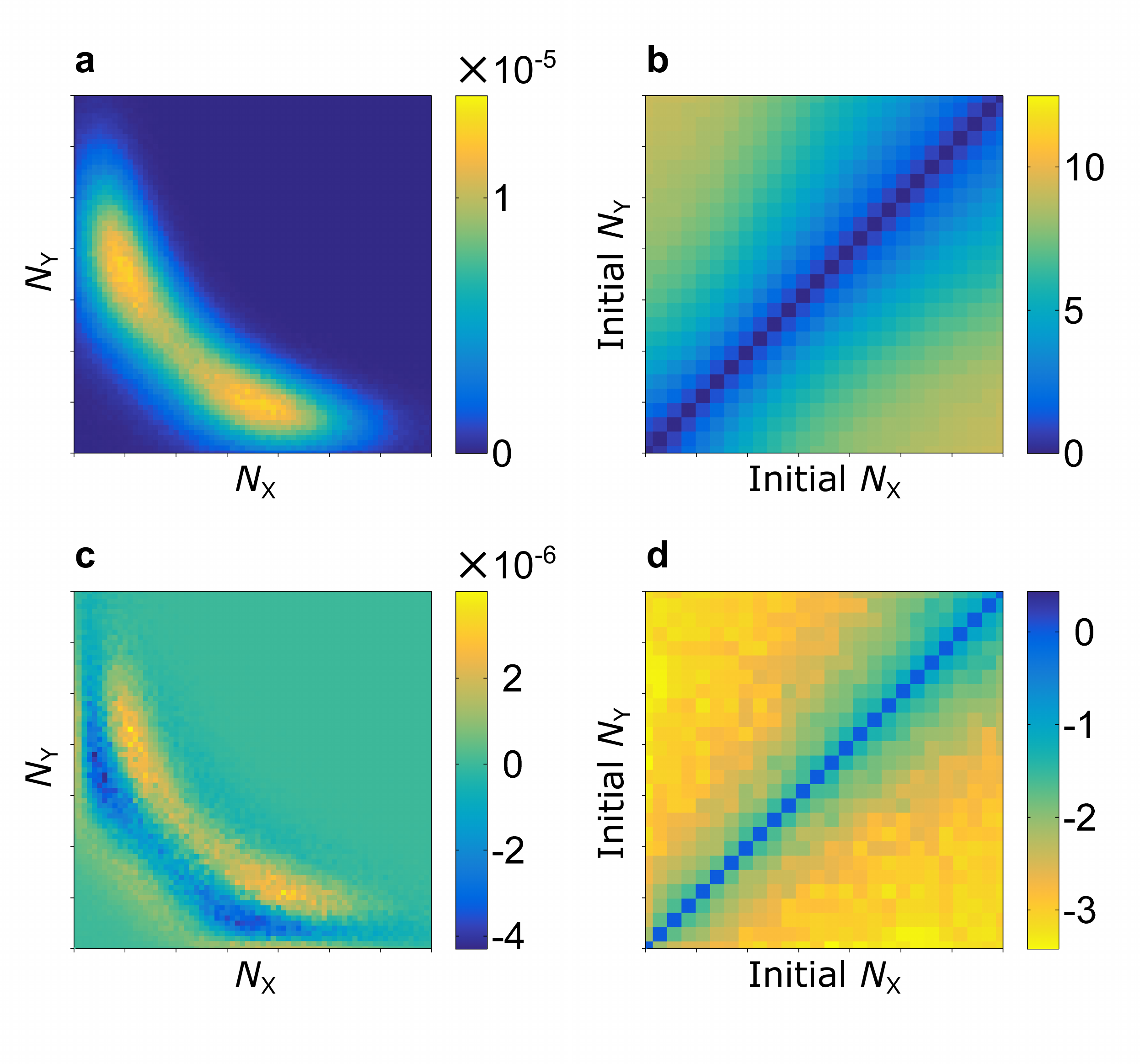}
\caption{Diffusion approximation of the protein-only model with geometrically distributed random bursts (GB); ({\bf a}) Stationary distribution as a function of the protein numbers; ({\bf b}) Mean first switching time (MFST) as a function of the initial protein numbers, in the unit of cell cycles; ({\bf c}) Net deviation of the stationary distribution from the full model; ({\bf d}) Net deviation of the MFST from the FM. All axes show the range $0\leq \NPI,\NPII\leq 700$ on linear scales. \label{fig:4}}
\end{center}
\end{figure}

Results from simulating the Gaussian process of equations \eqref{eq:diffapprox} are shown in Fig.~\ref{fig:4}. 
While the data for the stationary distribution (Fig.~\ref{fig:4}a) looks similar to that of the full model (Fig.~\ref{fig:2}a), noticeable discrepancies are manifest in the mean first switching times (compare Fig.~\ref{fig:4}b and Fig.~\ref{fig:3}a). 
In  Fig.~\ref{fig:4}c and d, we show the differences between simulation outcomes of the full model and those of the diffusion approximation of the GB model. Although the GB model itself approximates the full model well (Figs.~\ref{fig:2} and \ref{fig:3}), we conclude that the diffusion approximation fails to capture the relevant model outcomes.
\\

\noindent{\bf Constructing a mesoscopic piecewise deterministic Markov process.}
We have seen that the diffusion approximation of the GB model fails to reproduce the statistics of the full model. This underlines the need to construct coarse-grained models {\em directly from the full model} and without the intermediate step of a protein-only dynamics. We now proceed to introduce such a model. As before we describe protein concentrations by continuous variables, $x$ and $y$.  The mRNA dynamics are captured by introducing three `states': The $0$-state describes phases in which no mRNA is present. In the X-state there is one mRNA of type X and protein X is generated with rate $\gamma b$. The quantity $b=B/K$ is the mean burst size in the unit of protein concentration. No proteins of type Y are produced in the X-state. Similarly, in the Y-state protein Y is generated with rate $\gamma b$. Both types of protein are subject to natural degradation with rate $\gamma_0$ in any of the three states.

This is described by the following \emph{deterministic} differential equations:
\subeq{
\text{0-state:} {}& \quad \dot{x} = -\gamma_0 x \quad \text{and} \quad \dot{y}=-\gamma_0 y,\\
\text{X-state:} {}& \quad \dot{x} = \gamma b -\gamma_0 x \quad \text{and} \quad \dot{y}=-\gamma_0 y,\\
\text{Y-state:} {}& \quad \dot{x} = -\gamma_0 x \quad \text{and} \quad \dot{y}= \gamma b-\gamma_0 y.
}{}
The rates with which the system transits between the states are based on the dynamics of the FM:
\al{
\text{0-state} \xrightarrow{\HP\l(K y\r)} {}&\text{X-state, }~~~~~\text{X-state} \xrightarrow{\gamma}\text{0-state},\nonumber\\
\text{0-state} \xrightarrow{\HP\l(K x\r)} {}&\text{Y-state, }~~~~~\text{Y-state} \xrightarrow{\gamma}\text{0-state}.
}{eq:PDMP transition}
No transitions occur directly between the X and Y states. The kinetic scheme is illustrated in Fig.~\ref{fig:1}d. 

The stochasticity and discreteness of the mRNA populations is reflected in the random transitioning between the 0-, X- and Y-states. Between these Markovian events the protein concentrations evolve deterministically. We will refer to this model as the piecewise deterministic Markov process ({\bf PDMP}).

Notably, at most one mRNA molecule of either type can be present in PDMP at any time.  Although the model can be generalised to allow more than one mRNA molecule, the analysis below shows that the lowest-order approximation is sufficient to capture the relevant fluctuations of the mRNA dynamics. 
\\

\begin{figure}
\begin{center}
\includegraphics[width=0.48\textwidth]{\figureroot 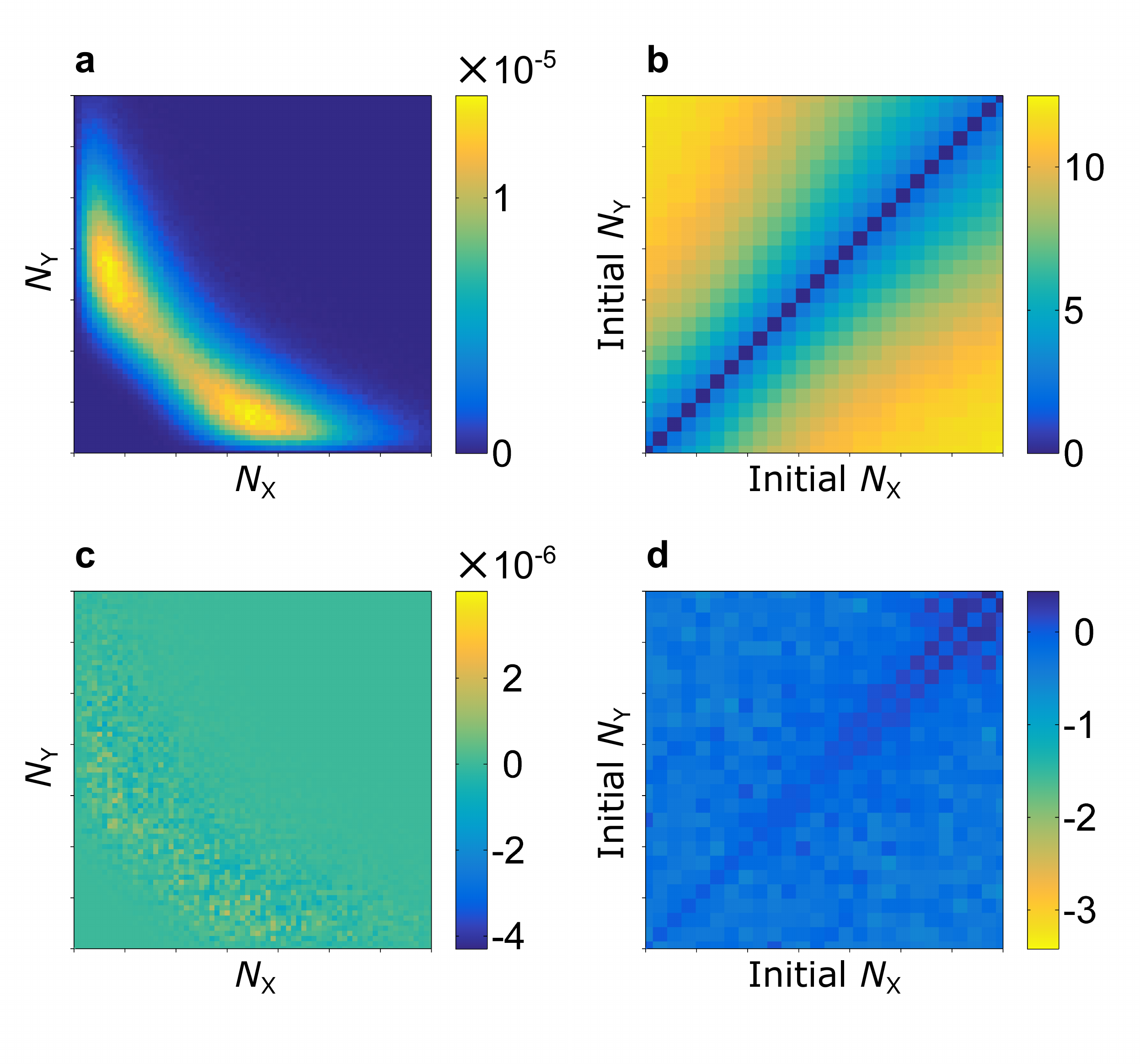}
\caption{PDMP approximation. ({\bf a}) Stationary distribution; ({\bf b}) Mean first switching time in the unit of cell cycles as a function of initial protein numbers. ({\bf c}) Net deviation of the stationary distribution from the full model; ({\bf d}) Net deviation of the MFST of the PDMP model from the FM. All axes are on linear scales and show the range $0\leq \NPI,\NPII\leq 700$. \label{fig:5}}
\end{center}
\end{figure}

\noindent{\bf The PDMP approximation outperforms the diffusion approximation of the GB model.}
As in the GB model, we work in the limit of infinitely fast degrading mRNA ($\gamma \rightarrow \infty$).  
Simulations of the PDMP model in this limit can be carried out using a minor modification of a previously proposed algorithm\cite{Bokes}.
We measure the stationary distribution of the PDMP model and the mean first switching times for different initial protein numbers. Results are shown in Fig.~\ref{fig:5}a and b, and we compare the outcome against that of the full model in Fig.~\ref{fig:5}c and d. 

The simulation data indicate that the PDMP approximation outperforms the diffusion approximation of the GB model, and it provides a more faithful approximation to the FM. This is because the diffusion approximation introduces Gaussian noise. It retains some information about the variance of protein production and degradation, but it does not capture the geometrically distributed burst sizes in the GB model well enough. The PDMP approximation, on the other hand, models exponentially distributed bursts in protein concentration. The exponential distribution in the PDMP model is the analogue of the geometric distribution in the discrete-molecule GB model. While the PDMP model is an approximation as well, it retains the typical characteristics of the stationary distribution and switching times of the original model.  At the same time the PDMP model is suitable for further mathematical analysis (see below). 
\\

\begin{figure}
\begin{center}
\includegraphics[width=0.48\textwidth]{\figureroot 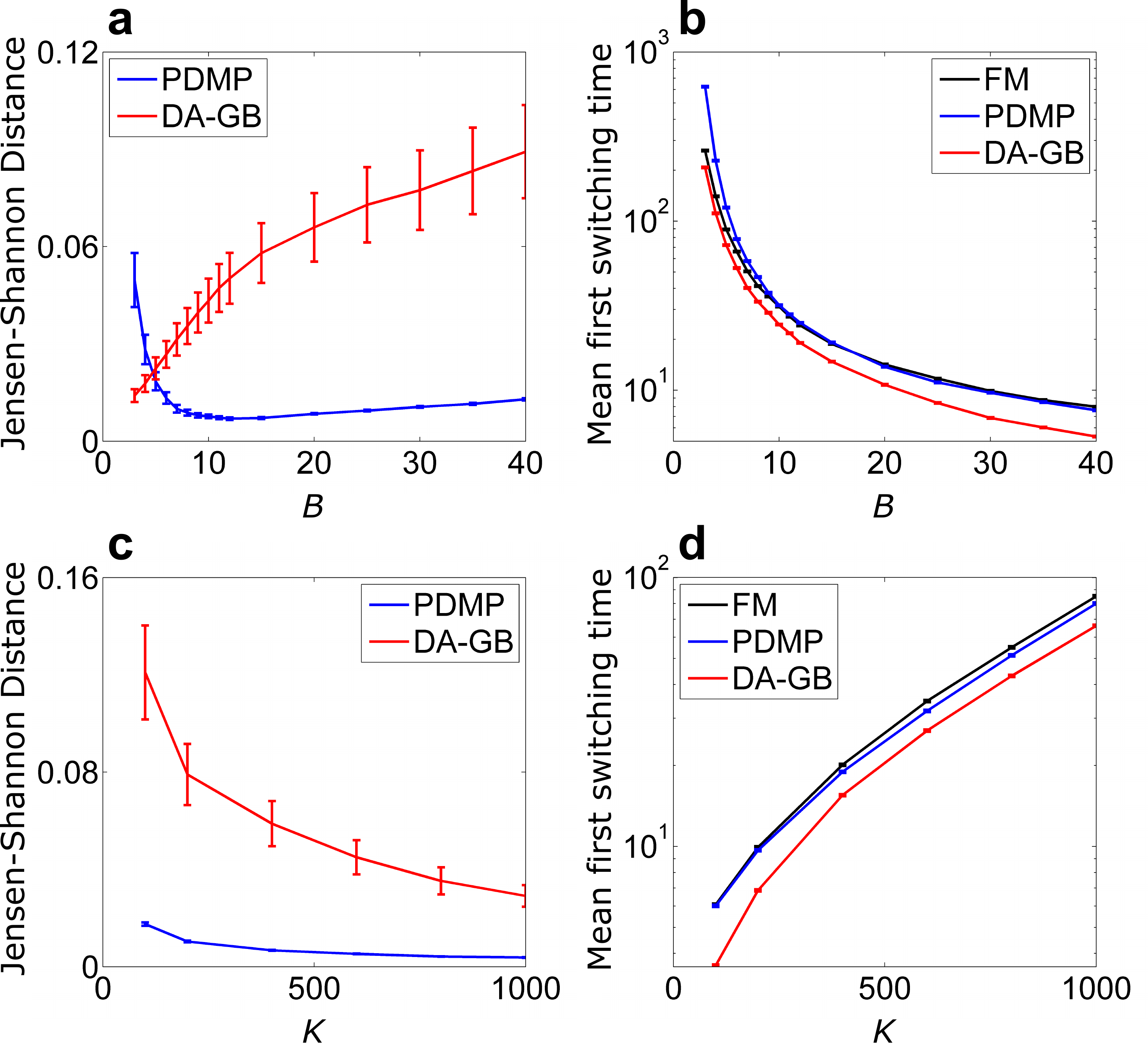}
\caption{Performance of the PDMP model and the diffusion approximation of the GB model (DA-GB). ({\bf a}) Jensen--Shanon distance between the stationary distribution PDMP (and DA-GB) and the stationary distribution of the FM; ({\bf b}) Mean first switching time for varying value of $B$ at fixed $K=200$; ({\bf c}-{\bf d}) Similar to ({\bf a}-{\bf b}) but now varying $K$ at fixed $B=30$. \label{fig:6}} 
\end{center}
\end{figure}

\noindent{\bf When does the PDMP outperform the diffusion approximation?}
We now investigate the robustness of these findings. In Fig.~\ref{fig:6} we vary two essential parameters, the mean burst size $B$, and the population scale $K$, while keeping the other parameters fixed.  We measure the Jensen--Shannon distance \cite{Lin,Endres} between the resulting stationary distributions of the PDMP and that of the the full model. 
Data is shown in Fig.~\ref{fig:6}a and c.  We also compare the mean first switching times starting from one of the stable modes, see Fig.~\ref{fig:6}b and d. The figure also shows results from the diffusion approximation of the GB model.

Results indicate that PDMP model outperforms the diffusion approximation of the GB model for mean burst sizes of $B\gtrsim 5$.  We conclude that the bursting noise has to be considered in this biologically relevant regime \cite{Swain}.  The PDMP model incorporates only the bursting noise and neglects the demographic noise from random degradation of the proteins. The strength of this demographic noise is proportional to $1/\sqrt{K}$. The results in Fig.~\ref{fig:6} c and d indicate that the difference in describing intrinsic noise propagates to physical observables even when the noise is weak ($K\approx 1000$ for fixed $B=30$).
\\

\noindent{\bf Analytic investigation of the PDMP process.}
The simplicity of the PDMP approach allows us to proceed with a mathematical analysis. We here only outline the main steps, further details are reported in the Supplementary Information.  We denote the probability density that the system is in the $0$-state and with protein densities $x,y$ at time $t$ by $\po(x,y,t)$. Similarly we write $\pX(x,y,t)$ and $\pY(x,y,t)$ when the system is in the X- or Y-states. The evolution of these distributions then follows the forward equation
\eq{
\frac{\partial}{\partial t} \l[
\begin{array}{c}
\po\\\pX\\\pY
\end{array}
\r]
= \l(L^{\dagger}_{d} + L^{\dagger}_\text{s}\r) \l[
\begin{array}{c}
\po\\\pX\\\pY
\end{array}
\r],
}{eq:PDMP}
where $L^{\dagger}_d$ and $L^{\dagger}_s$ drive the deterministic flow and the random switching between states respectively. These operators are of the form
\subeq{
 L^{\dagger}_d :={}& \l[
\begin{array}{ccc}
\l(L^\dagger_d\r)_{11} & 0 & 0\\
0 &\l(L^\dagger_d\r)_{22} & 0 \\
0&0 & \l(L^\dagger_d\r)_{33}
\end{array}
\r], \\
 L^{\dagger}_s :={}& \l[
\begin{array}{ccc}
- \HP(Kx)-\HP(Ky)  & \gamma & \gamma\\
\HP(Ky) &- \gamma & 0 \\
 \HP(Kx)&0 & - \gamma
\end{array}
\r],
}{eq:PDMP operators}
with
\subeq{
\l(L^\dagger_d\r)_{11} :={}&\gamma_0\partial_x \l(x\r) + \gamma_0\partial_y \l(y\r),\\
\l(L^\dagger_d\r)_{22} :={}& \partial_x \l(-\gamma b + \gamma_0 x\r) + \gamma_0 \partial_y \l(y\r) ,\\
\l(L^\dagger_d\r)_{33} :={}&\gamma_0 \partial_x \l(x\r) + \partial_y \l(-\gamma b + \gamma_0y\r).
}{}
The differential operators $\partial_x$ and $\partial_y$ act on all that follows to their right, including the probability densities $p_0, \pX$ and $\pY$ outside the matrix notation in equations \eqref{eq:PDMP} and \eqref{eq:PDMP operators}.

The PDMP approximation applies in the limit $\gamma\rightarrow \infty$, i.e., fast return into the 0-state.  The resident time in the X- and Y-states is exponentially distributed and scales as $\gamma^{-1}$. It formally tends to zero as $\gamma\to\infty$. 
On the other hand the translation rate $\gamma B$ tends to infinity in this limit. 
Combining the limiting behaviours of resident time and translation rate results in an \emph{exponentially distributed} increment of protein concentration in each cycle of switching from the 0-state to the X- or Y-state, and then returning to the 0-state. 
As a consequence the PDMP converges to previously proposed continuous-state bursting models \cite{Friedman,Bokes,Cai} in the limit $\gamma \rightarrow \infty$, and $\po(x,y,t)$ satisfies 
\al{
\partial_t p_0={}& \partial_x \l( \gamma_0 x p_0 \r) + \partial_y \l( \gamma_0 y p_0\r) - \l[ \HP(Kx) + \HP(Ky) \r] p_0 \nonumber\\
{}&+ \HP(Ky) \int_0^{x} \frac{1}{b} e^{-\frac{x-x'}{b}} p_0\l(x',y,t\r)dx'  \nonumber\\
{}&+  \HP(Kx) \int_0^{y} \frac{1}{b} e^{-\frac{y-y'}{b}} p_0\l(x,y',t\r) dy',
}{eq:PDMP ID}
as detailed in the Supplementary Information.
\\

\noindent{\bf Analytic investigation of the mean first switching time.} 
One of the strengths of the PDMP formulation (equations \eqref{eq:PDMP} and \eqref{eq:PDMP operators}) is the relative ease with which mean first switching times can be obtained. We first proceed by computing mean escape time from an arbitrary open domain $\Omega$. The mean first switching time can be calculated by setting $\Omega=\l\{(x,y):x<y\r\}$, recognising that the process can only exit this domain by crossing the boundary $x=y$.

\begin{figure}
\begin{center}
\includegraphics[width=0.48\textwidth]{\figureroot 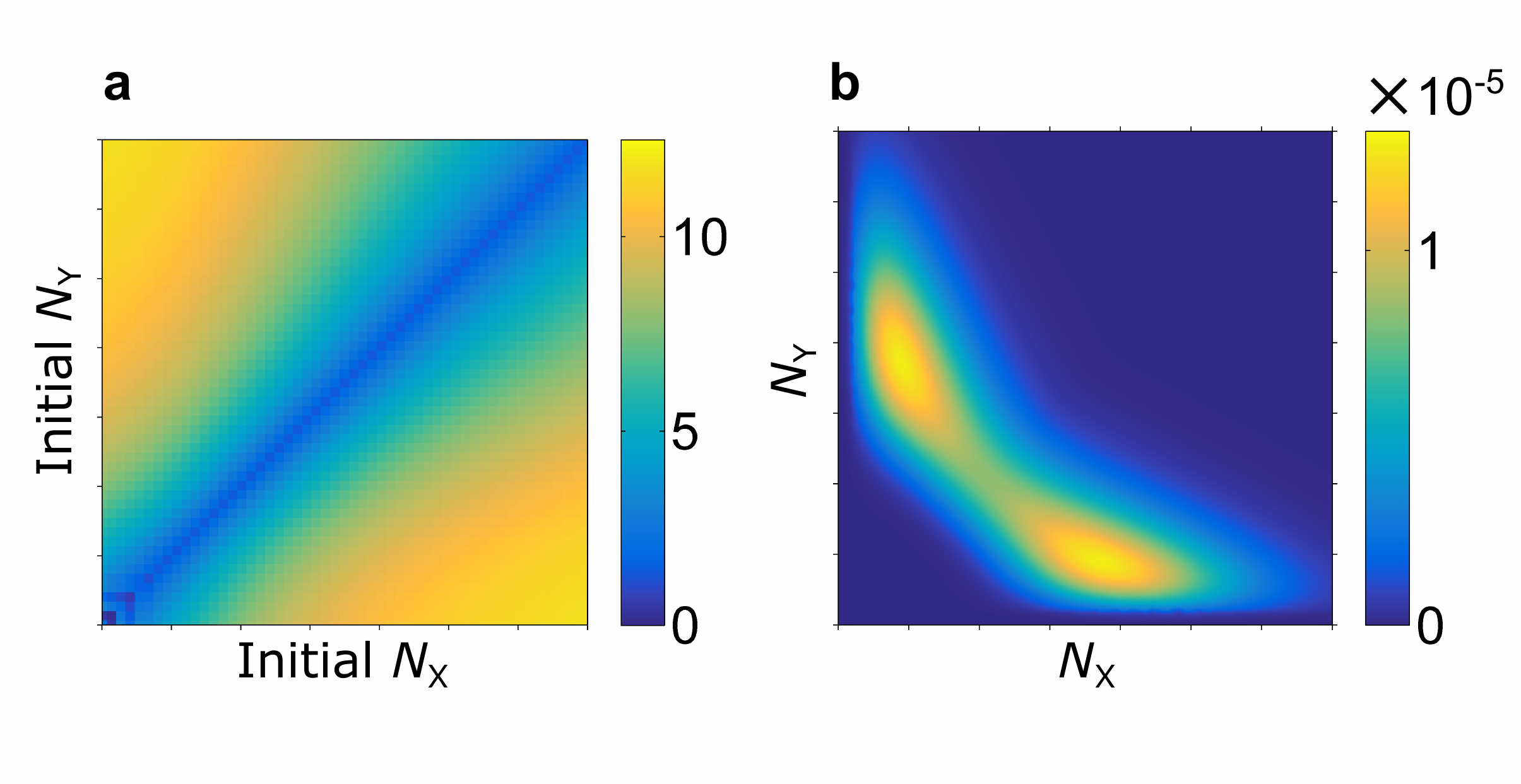}
\caption{Theoretical prediction of the PDMP model. ({\bf a}) Mean first passage time as a function of initial protein numbers, calculated from the backward equation \eqref{eq:integro-diff}; ({\bf b}) Stationary distribution of protein numbers calculated from the WKB method. Axes of both panels show the range $0\leq \NPI(0),\NPII(0)\leq 700$ on linear scales. \label{fig:7}}
\end{center}
\end{figure}

Suppose, the system is initially at $(x,y)\in \Omega$, and in state $\text{Z}\in \l\{0, \text{X},\text{Y}\r\}$. We write $T_\text{Z}\l(x,y\r)$ for the mean first time at which the process exits the domain $\Omega$. The quantities $T_\text{Z}$ then satisfy the following backward equation \cite{vanKampen,Gardiner}
\eq{
-\l[
\begin{array}{c}
1\\ 1\\ 1
\end{array}
\r]
= \l(L_{d} + L_\text{s}\r) \l[
\begin{array}{c}
T_0(x,y) \\T_\text{X}(x,y)\\T_\text{Y}(x,y)
\end{array}
\r],
}{eq:backward eq}
where $L_d$ and $L_s$ are adjoint to the operators in equations \eqref{eq:PDMP operators}. They are given by
\subeq{
 L_d :={}& \l[
\begin{array}{ccc}
\l(L_d\r)_{11} & 0 & 0\\
0 &\l(L_d\r)_{22} & 0 \\
0&0 & \l(L_d\r)_{33}
\end{array}
\r], \\
 L_s :={}& \l[
\begin{array}{ccc}
- \HP(Kx)-\HP(Ky)  & \HP(Ky) &  \HP(Kx)\\
 \gamma &- \gamma & 0 \\
\gamma &0 & - \gamma
\end{array}
\r]
}{eq:b-operators}
with
\subeq{
\l(L_d\r)_{11} ={}&-\gamma_0 x\partial_x  - \gamma_0 y\partial_y ,\\
\l(L_d\r)_{22} ={}&\l(\gamma b - \gamma_0 x\r) \partial_x  - \gamma_0 \l(y\r) \partial_y,\\
\l(L_d\r)_{33} ={}&-\gamma_0 x \partial_x +\l(\gamma b - \gamma_0y\r)  \partial_y .
}{}
In the infinitely-fast degrading mRNA limit ($\gamma\rightarrow \infty$), and using appropriate boundary conditions (Supplementary Information) we arrive at
\al{
-1 ={}& \l[-\gamma_0 x \partial_x  -\gamma_0 y\partial_y - H(Kx)-H(Ky)\r] T_0\l(x,y\r) \nonumber\\
{}&+ H(Ky) \int_x^y \frac{e^{-\frac{x'-x}{b}}}{b}T_0\l(x',y\r) dx' \nonumber\\
{}&+ H(Kx) \int_y^\infty \frac{e^{-\frac{y'-y}{b}}}{b} T_0\l(x,y'\r) dx'.
}{eq:integro-diff}
This is the adjoint equation\cite{vanKampen} of the expression in equation \eqref{eq:PDMP ID} on the open domain $\Omega$. 
Equation \eqref{eq:integro-diff} is solved by a finite difference method, noting that it is self-consistent and no boundary condition needs to be specified.  The solution is shown in Fig.~\ref{fig:7}, and reproduces the simulation outcome of the FM well. 

We remark that equation \eqref{eq:integro-diff} is only valid for the half-plane $\Omega$. A detailed discussion can be found in the Supplementary Information.
\\

\begin{figure}
\begin{center}
\includegraphics[width=0.48\textwidth]{\figureroot 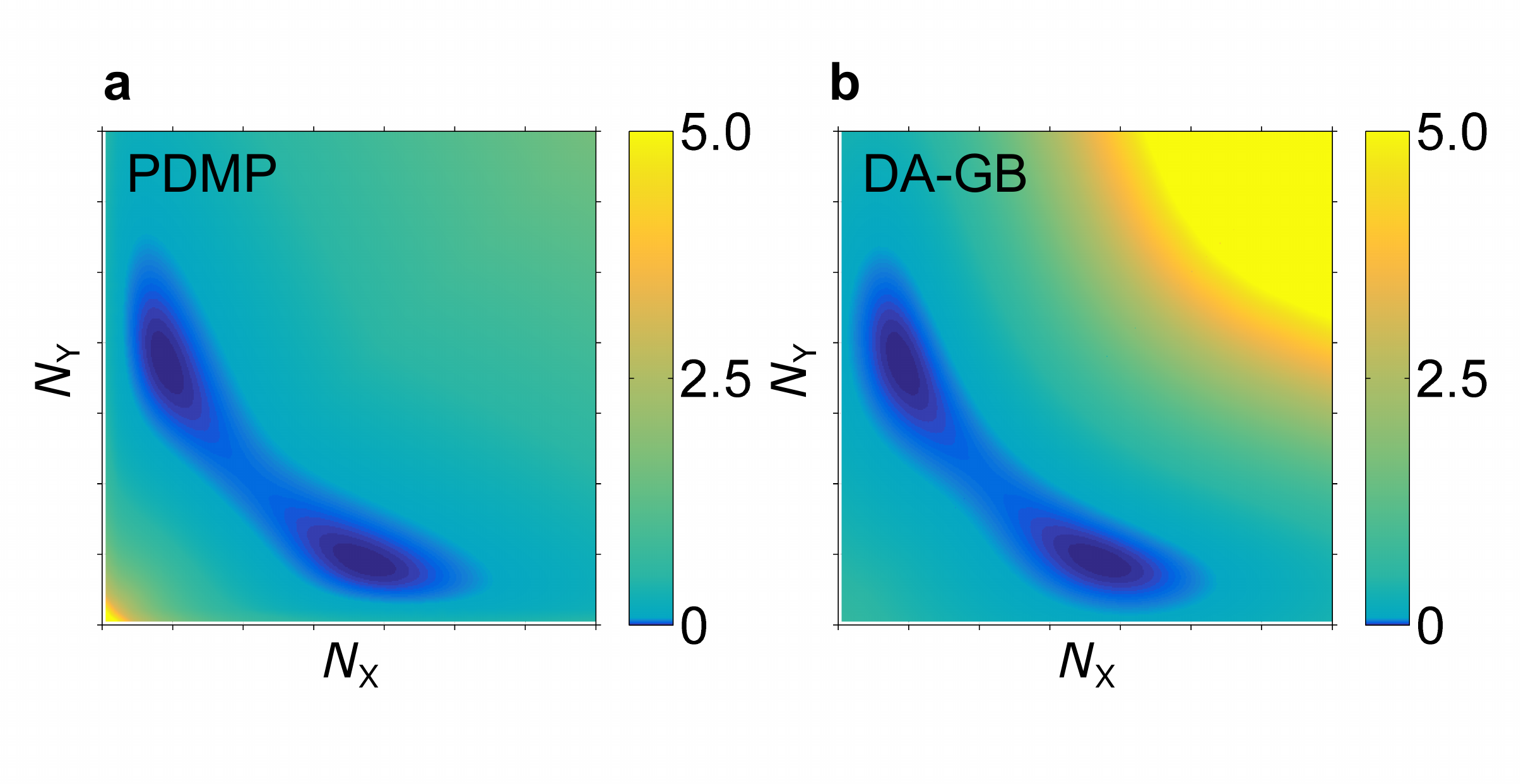}
\caption{Rate functions $S_0$ as functions of the protein numbers $0\leq \NPI,\NPII\leq 700$ on a linear scale.  ({\bf a}) PDMP model; ({\bf b}) Diffusion approximation of the GB model. \label{fig:8}}
\end{center}
\end{figure}

\noindent{\bf Analytic investigation of the weak-noise limit.}
The analytical calculation of the stationary distributions of the PDMP model can be pursued further using the so-called Wentzel-Kramers-Brillouin (WKB) method. This technique is based on the ansatz
\begin{equation}
p_{\rm stat}\l(x,y\r) =\exp \l[-\frac{1}{b} \sum_{\ell=0}^\infty b^\ell S_\ell\l(x, y\r)\r], 
\label{eq:WKB ansatz}
\end{equation}
where $b=B/K\ll 1$.  One proceeds by considering $\l(L_d^\dagger+ L_s^\dagger\r) p_{\rm stat}(x,y)=0$ order-by-order in $b$. To leading order we find the Hamilton--Jacobi equation
\al{
0 ={}& \l[\gamma_0 x - Bh\l(y\r)\r)] \partial_x S_0 + \l[\gamma_0 y - Bh\l(x\r)\r]  \partial_y S_0 \nonumber \\
{}& + \l[\gamma_0 x+\gamma_0 y -Bh\l(x\r) - Bh\l(y\r)\r] \l(\partial_x S_0\r)\l(\partial_y  S_0\r) \nonumber\\
{}&+\gamma_0 x \l(\partial_x S_0\r)^2 + \gamma_0 y \l(\partial_y S_0\r)^2  \nonumber \\
{}&+\gamma_0 x \l(\partial_x S_0\r)^2\l(\partial_y  S_0\r) +   \gamma_0 y \l(\partial_x S_0\r)\l(\partial_y  S_0\r)^2,
}{eq:PDMPHJ}
where $h(z):=H(Kz)/K$. 
This  equation is then numerically solved using the algorithm of Heymann and Vanden--Eijnden \cite{Heymann}. Results are shown in Fig.~\ref{fig:8}.  Even though this only provides a first-order approximation and despite the fact that we have used $b=0.15$ (which is not very small) we obtain a reasonable agreement with the stationary distribution in Fig.~\ref{fig:5}a. 

For completeness we have also carried out a WKB analysis of the diffusion approximation of the GB, CB and NB models. These are presented in the Supplementary Information.

The leading order function $S_0\l(x,y\r)$ is the so-called `rate function' which quantifies the rare-event statistics of the process in the weak-noise limit $b\ll 1$ \cite{WKB,Zhou}.  
Several studies have suggested that $S_0\l(x,y\r)$ is a suitable candidate for a `landscape' of the non-equilibrium random processes in models of gene regulatory networks\cite{Wang,Warren,Strasser,Assaf,Roma}. 
The Hamilton--Jacobi equation \eqref{eq:PDMPHJ} contains cubic terms such as $\l(\partial_x S_0\r)^2 \l(\partial_y S_0\r)$, while diffusion equations are quadratic in derivatives of $S_0$. This illustrates the fundamental difference between the statistics of intrinsic noise in the diffusion approximation and the bursting noise in PDMP. Further more rigorous mathematical investigations into these differences would be very welcome in our view.

We compare the functions $S_0$ of the PDMP and the diffusion approximation of the GB model in Fig.~\ref{fig:8}. One observes a much `shallower' rate function in the PMDP model, especially at larger protein numbers ($\NPI,\NPII\approx 700$).  This is due to the long tails in the exponential bursting kernel of the PDMP model, which are not present in the diffusion approximation of the GB model. Such a fat-tail bursting kernel enhances the probability for the system to evolve to high protein concentrations. We identify this as the origin of the qualitatively distinct rare-event statistics in the two models. 
\\

\begin{figure}
\begin{center}
\includegraphics[width=0.48\textwidth]{\figureroot 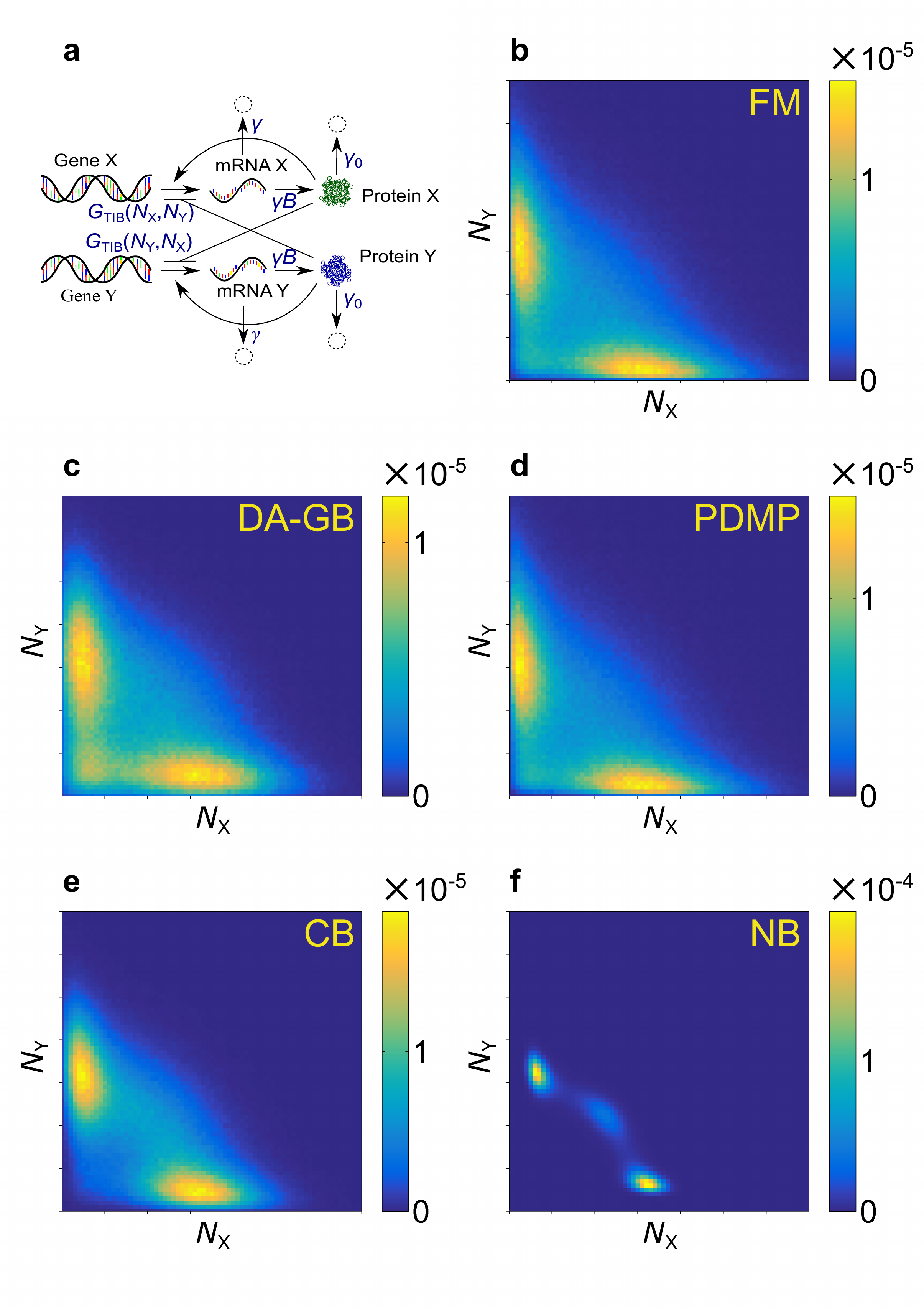}
\caption{({\bf a}) Schematic diagram illustrating the network of the three-way switch, remaining panels show stationary distribution of protein numbers in the range $0\leq N_X,N_Y\leq 700$ on linear scale. ({\bf b})  Full model;  ({\bf c}) Diffusion approximation of the GB model; ({\bf d}) PDMP approximation;  ({\bf e}) CB model; and ({\bf f})NB model. \label{fig:9}}
\end{center}
\end{figure}

\noindent{\bf Effects of bursting noise in a multi-switch network.}
Recently, multi-switch systems have gained interest\cite{Lu,Lu2,Guantes}. 
A schematic diagram of the three-way switch network proposed by \cite{Lu} is shown in Fig.~\ref{fig:9}a. 
It is obtained from the classical toggle switch network by including a self-enhancing autoregulation. 
Our computational and mathematical setup requires only minor modifications to include generalisation to this case. 
Specifically, we replace the earlier Hill functions by 
\al{
G(\NPI,\NPII) ={}& q_0 \l(1 + \frac{r_1 }{(\NPI/K_1)^{n_1} + 1}\r)\nonumber \\
{}&\times \l(1 + \frac{r_2 }{(\NPII/K_2)^{n_2} + 1}\r),
}{}
with parameters\cite{Lu} $q_0= 4$, $r_1= -4/5$, $r_2 = 7/3$, $n_1=3$, $n_2 =1$, $K_1=160$, and $K_2=320$. The rest of the parameters follows Table \ref{table:1}. The negative value of $r_1$ reflects the positive autoregulation. To evaluate the effects of bursting noise on this multi-switch model, we consider again the full model, the diffusion approximation of the GB model, as well as the CB and NB models of the extended network. 

Fig.~\ref{fig:9} displays the stationary distribution to illustrate the effects of the bursting noise in the multi-switch network. 
The model without bursts (NB, panel f) has a stationary distribution consisting of three modes, as reported earlier\cite{Lu}. Inclusion of constant bursts (CB, panel e) diversifies the protein expression and reduces the stability of the mode located at $\NPI=\NPII\approx 230$. In the full model (panel b) there is no discernible concentration of probability in the symmetric mode, hence the three-way switching capability appears to be absent. 
We also notice that the saddle of the distribution in the FM is located at a state with much lower number of proteins compared to the NB and CB models. 
The most likely switching path \cite{Roma} from one of the asymmetric modes to the other will differ significantly between the different variants of the model.  
The diffusion approximation of the GB model (panel c) does not capture the outcome of the FM either. 
Overall these findings confirm again that the inclusion of bursty noise statistics has significant effects on the model outcome.  
Finally, we observe in Fig.~\ref{fig:9}d that the PDMP model approximates the full model of the three-way switch well. We conclude that randomly distributed burst sizes are again the predominant form of intrinsic noise in the multi-switch network.

\section{Discussion and conclusion}

Explicitly including mRNA dynamics in gene regulatory models inevitably introduces more complexity. We have quantitatively studied the effects of bursting noise \cite{Kaern} in a biologically relevant regime or the model organism \emph{E.~coli}.  To our knowledge, this is one of the first which attempts to build a rigorous connection between existing individual-based models\cite{Walczak,WalczakSasai,Roma,Warren} and more coarse-grained models\cite{Wang,WangHuang,Friedman,Bokes}. 
Results of our simulations indicate that the bursting statistics of transcription and translation are essential ingredients of models of gene regulation. Coarse grained models need to account for bursting to retain correct statistics of noise-driven phenomena such as the switching between different dynamic attractors.

The implications of our observations are relevant to the abstract modelling of regulatory networks in different ways. 
We are now in a better position to address our opening question, and to say how noise propagates between different levels of modelling. Perhaps more importantly, our study may ultimately help to decide what level of modelling is most appropriate to study gene regulatory circuits computationally. The answer will of course depend on the question in the focus of the investigation. We have examined different levels of coarse graining, and we have identified the steps in these reduction procedures at which significant alternations to different model outcomes are introduced. 

Systematically choosing a suitable level of coarse-graining also facilitates the mathematical analysis of regulatory networks.  
The high dimensionality of full regulatory network effectively makes them intractable. Model reduction is needed  to make progress, and our analysis demonstrates that the PDMP formulation is a powerful way forward, and that it can be more suitable than the conventional diffusion approximation. The PDMP model explicitly retains the bursting noise originating from the mRNA dynamics. Even though it effectively disregards the demographic noise from random degradation of the proteins, it delivers accurate predictions for stationary distributions and switching times. 

As another strength, the PDMP formulation can relatively easily be generalised to accommodate more complex reactions. For example, in the \emph{Enterobacteria phage $\lambda$} switch it is not the monomer of the synthesized proteins which acts as the repressor to regulate transcription, but instead their dimer. Modelling these processes requires the inclusion of dimerization further downstream after transcription and translation \cite{Arkin,Warren}. Preliminary results not shown here reveal that the PDMP approximates such dynamics well. 

The fact that the piecewise deterministic Markov process is successful in approximating the full model opens a relatively new type of modelling paradigm. We acknowledge that we are not the first to propose this \cite{Hasty1,Zeiser,Zeiser2,Hu,Kumar}. 
Our contribution consists in a first analytical treatment of PDMP models and in an systematic embedding into a wider landscape of modelling approaches. The bursting phenomenon is ubiquitous whenever there is a separation of time scales between the source  and the product of a biological process. These are mRNA and protein in models of gene regulation, but we expect that these ideas can be applied to other biological problems with similar time-scale separation.

\section{Methods}
Sample paths of the individual-based processes (FM, CB, NB, and GB) are generated using the standard Gillespie algorithm \cite{Schwartz,Gillespie}. The PDMP process is simulated using the algorithm discussed by Bokes et al\cite{Bokes}. Simulations of the stochastic differential equations resulting from the diffusion approximation are performed with a standard Euler--Maruyama scheme. The geometric minimum action method\cite{Heymann} is implemented using MATLAB R2010a, as is the finite-difference scheme to solve the backward equation \eqref{eq:integro-diff}. Further details can be found in the Supplementary Information.

\section{Acknowledgement}
The authors acknowledge supported by the Engineering and Physical Sciences Research Council EPSRC (UK) (grant reference EP/K037145/1). We thank Peter Ashcroft, Luis Fernandez--Lafuerza, Louise Dyson, and Charlie Doering for fruitful discussions. 

\section{Author contributions}
Y.T.L.~conceived the study, performed analytical calculations and numerical simulations, prepared the figures. T.G.~provided general direction of the study, confirming analytical calculations. Both authors wrote the manuscript.

\section{Additional information}
\noindent{\bf Supplementary Information} accompanies this manuscript.

\noindent{\bf Competing financial interests:} The authors declare no competing financial interests.

\end{document}

%\preprint{APS/123-QED}

\title{Supplementary Information\\Bursting noise in gene expression dynamics: \\ Linking microscopic and mesoscopic models}

\author{Yen Ting Lin}
\email{yentingl@umich.edu}
\affiliation{Theoretical Physics, School of Physics and Astronomy, The University of 
Manchester, Manchester M13 9PL, UK}

\author{Tobias Galla}
\email{tobias.galla@manchester.ac.uk}
\affiliation{Theoretical Physics, School of Physics and Astronomy, The University of  Manchester, Manchester M13 9PL, UK}

\date{\today}

\maketitle

\tableofcontents

\newcommand{\NPI}{N_{\rm X}}
\newcommand{\NPII}{N_{\rm Y}}
\newcommand{\NMI}{M_{\rm X}}
\newcommand{\NMII}{M_{\rm Y}}
\newcommand{\NPIII}{N_{\rm Z}}
\newcommand{\HMF}{\frac{B}{K}H}
\newcommand{\HSIBB}{H_{}}
\newcommand{\HSIB}{B H_{}}
\newcommand{\HIB}{H_{}}
\newcommand{\HP}{H_{}}
\newcommand{\po}{p_\text{0}}
\newcommand{\pX}{p_\text{X}}
\newcommand{\pY}{p_\text{Y}}
\newcommand{\GIB}{G_\text{TIB}}
\newcommand{\TIB}{TIB }
\newcommand{\SIBB}{SIBB }
\newcommand{\SIB}{SIB }
%\setcitestyle{super}
%\renewcommand*{\bibnumfmt}[1]{#1}

\newpage

\section{Notation}
We briefly summarise the notation used in the main manuscript and in this supplement:
\begin{itemize}
\item  Discrete numbers of the two types of protein are denoted by $\NPI$ and $\NPII$. We write $\NMI$ and $\NMII$ for the number of mRNA molecules of the two types.
\item Variables such as $x(t)$ denote continuous particle densities (or concentrations). Specifically $x(t)$ and $y(t)$ are protein densities, i.e., $x=\NPI/K$ and $y=\NPII/K$ in the limit $K\gg 1$.
\item We denote the probabilities in the master equations by capital $P$ (discrete particle numbers).
\item The lower-case notation $p$ is used for probability density functions in the diffusion approximation (continuous particle densities/concentrations).

\end{itemize}

\section{Master equations of the individual-based models}
The different individual-based model in the main manuscript are uniquely defined by their master equations. Here we briefly summarise the master equations of the FM and of the GB, CB and NB models.

\subsection{Master equation of full model (FM)}
We write $P_{a,b,c,d}$ for the probability that the system is in state $\NMI=a,\NMII=b,\NPI=c,\NPII=d$ at time $t$. The master equation of the FM is then
\begin{align}
\frac{d}{dt} P_{a,b,c,d} ={}& - \l\{\HIB\l(c\r)+\HIB\l(d\r) + a\gamma \l[ 1+B \r] + b\gamma\l[1+B\r] - \gamma_0 c - \gamma_0 d\r\} P_{a,b,c,d} \nonumber \\
							   {}&+ \HIB\l(c\r) P_{a, b-1, c, d} + \HIB\l(d\r) P_{a-1, b, c, d} + \gamma \l(a+1\r)  P_{a+1, b, c, d}+  \gamma \l(b+1\r) P_{a, b+1, c, d}\nonumber \\
							   {}&+ B\gamma a P_{a,b,c-1,d} + B\gamma bP_{a, b,c,d-1} + \gamma_0 \l(c+1\r) P_{a,b,c+1,d}+ \gamma_0 \l(d+1\r) P_{a,b,c,d+1}.  \label{eq:masterTSIB}
\end{align}
The probability of a state is zero if any of the variables $a,b,c$ or $d$ are negative.

\subsection{Infinitely fast-degrading mRNA limit: the GB model}
In the kinetic scheme of the full model the mRNA decays with a rate $\gamma$, but synthesizes a protein with a rate $\gamma B$. Both of the rates are constants. One an mRNA is created the next event involving this mRNA particle is either the production of protein or the decay of the mRNA molecule. The probability that the next event is the synthesis of a protein is $B/(B+1)$, and the probability that a decay occurs next (before production of a protein) is $1/(B+1)$. The random number, $\ell$, of protein molecules generated by one particular mRNA molecule during its lifetime then follows a geometric distribution
\eq{
g(\ell)= \l(\frac{B}{1+B}\r)^\ell \l(\frac{1}{1+B}\r).
}{eq:geo}
The lifetime of an mRNA molecule is of order $\mathcal{O}\l(1/\gamma\r)$. As a consequence, we can think of the protein-generating process as follows in the infinitely-fast decaying mRNA limit ($\gamma \rightarrow \infty$): As soon as an mRNA is transcribed, it immediately releases a random number of proteins $\ell$ drawn from the distribution \eqref{eq:geo} and then decays.

\subsection{Master equation of the GB model}
In the limit $\gamma \rightarrow \infty$, the dynamics of the full model can effectively be coarse-grained into a single-species model
\subeq{
\text{Gene X} \xrightarrow{\HIB\l(\NPII\r)} {}&  \ell \times \text{Protein X} \quad \text{(transcription and translation of protein X)} , \\
\text{Gene Y} \xrightarrow{\HIB\l(\NPI\r)}  {}& \ell \times \text{Protein Y} \quad \text{(transcription and translation of protein Y)},\\
\text{Protein X} \xrightarrow{\gamma_0}  {}&  \emptyset \quad \quad \text{(degradation of protein X)}, \\
\text{Protein Y} \xrightarrow{\gamma_0}  {}&  \emptyset \quad \quad \text{(degradation of protein Y)},
}{eq:geo IB}
where $\ell$ is drawn from the above geometric distribution, every time one of the first two reactions fires.
The master equation of this process is
\al{
\frac{d}{dt} P_{c,d} ={}& -\l[\HIB\l(c\r) + \HIB\l(d\r) + \gamma_0 c + \gamma_0 d \r] P_{c,d} +  \gamma_0 \l(c+1\r) P_{c+1,d} +  \gamma_0 \l(d+1\r) P_{c,d+1}\nonumber \\
							{}&+ \sum_{\ell=0}^{c} \HIB\l(d\r)  \l(\frac{B}{1+B}\r)^\ell \l(\frac{1}{1+B}\r)  P_{c-\ell,d} + \sum_{\ell=0}^{d} \HIB\l(c\r)  \l(\frac{B}{1+B}\r)^\ell \l(\frac{1}{1+B}\r)  P_{c,d-\ell}.
}{eq:master geo}
We have written $P_{c,d}(t)$ for the probability that the system is in state $\NPI=c,\NPII=d$ at time $t$.
Again, the probability of a state is zero if $c$ or $d$ are negative.

\subsection{Master equation of CB model}
The master equation for the CB model is obtained by replacing $g(\ell)\rightarrow \delta_{\ell,B}$, i.e., $\ell$ takes value $\ell=B$ with probability one. We find
\begin{align}
\frac{d}{dt} P_{c,d} ={}& -\l[\HSIBB\l(c\r) + \HSIBB\l(d\r) + \gamma_0 c + \gamma_0 d \r] P_{c,d} \nonumber \\
							{}&+ \HSIBB\l(d\r) P_{c-B,d} + \HSIBB\l(c\r) P_{c, d-B} +  \gamma_0 \l(c+1\r) P_{c+1,d} +  \gamma_0 \l(d+1\r) P_{c,d+1}.
\end{align}

\subsection{Master equation of the model without bursts (NB)}
In this case we have
\begin{align}
\frac{d}{dt} P_{c,d} ={}& -\l[\HSIB\l(c\r) + \HSIB\l(d\r) + \gamma_0 c + \gamma_0 d \r] P_{c,d} \nonumber \\
							{}&+ \HSIB\l(d\r) P_{c-1,d} + \HSIB\l(c\r) P_{c, d-1} +  \gamma_0 \l(c+1\r) P_{c+1,d} +  \gamma_0 \l(d+1\r) P_{c,d+1}.
\end{align}

\section{Deriving the diffusion approximations}
\subsection{Diffusion approximation of the GB model}
Simulations of the full model (FM) show that the number of mRNA molecules present at any one time is typically very small ($\NMI,\NMII<10$) when biologically relevant parameters are used. The conventional diffusion approximation relies on large particle numbers, and so it is not adequate for the full model. 

Instead, we perform the diffusion approximation to the master equation of the GB model, equation \eqref{eq:master geo}. This is a standard method, and we proceed along the lines of \cite{vanKampen}. The only complication is the presence of the geometrically distributed random numbers (denoted by $\ell$) in the protein-generation reactions. As we will discuss below this requires only modest modifications to the standard Kramers-Moyal expansion.

We assume that the scale of the population size, $K$, is large but finite, i.e., $K\gg 1$, and we write $x =\NPI/K$ and $y = \NPII/K$, and replace $P_{c,d}(t)$ in favour of $p(x,y,t)$. 
 The master equation \eqref{eq:master geo} then becomes  
\al{\partial_t p(x,y,t) = {}& -\l[ \HIB\l(Kx\r) + \HIB\l(Ky\r) + \gamma_0 Kx + \gamma_0 Ky \r] p\l(x,y,t\r) \nonumber \\
						{}&+  \gamma_0 K \l(x+\frac{1}{K}\r) p\l(x+\frac{1}{K},y,t\r) +   \gamma_0 K \l(y+\frac{1}{K}\r) p\l(x, y+\frac{1}{K},t\r) \nonumber \\
						{}& + \sum_{\ell=0}^{\infty} \HIB\l(Ky\r)  \l(\frac{B}{1+B}\r)^\ell \l(\frac{1}{1+B}\r)  p\l(x-\frac{\ell}{K},y,t\r) \nonumber \\
						{}& + \sum_{\ell=0}^{\infty} \HIB\l(Kx\r)  \l(\frac{B}{1+B}\r)^\ell \l(\frac{1}{1+B}\r)  p\l(x,y-\frac{\ell}{K},t\r).
}{}
In the last two terms we have extended the summation over $\ell$ to infinity, terms in which $x-\ell/K$ or $y-\ell/K$ become negative are automatically suppressed as the corresponding probabilities $p(x-\ell/K,y,t)$ and $p(x,y-\ell/K,t)$ vanish.

The above expression can then be written as
\al{
\partial_t p(x,y,t) = {}& -\l[ \HIB\l(Kx\r) + \HIB\l(Ky\r) + \gamma_0 Kx + \gamma_0 Ky \r] p\l(x,y,t\r) \nonumber \\
						{}&+  \gamma_0 K \l(x+\frac{1}{K}\r) p\l(x+\frac{1}{K},y,t\r) +   \gamma_0 K \l(y+\frac{1}{K}\r) p\l(x, y+\frac{1}{K},t\r) \nonumber \\
						{}& +\HIB\l(Ky\r)\l\langle   p\l(x-\frac{\ell}{K},y,t\r) \r\rangle_\ell + \HIB\l(Kx\r) \l\langle   p\l(x,y-\frac{\ell}{K},t\r) \r\rangle_\ell,
}{}
where $\langle \cdots \rangle_l$ denotes an average with respect to a geometrically distributed random number $\ell$, i.e., $\langle f_\ell\rangle_\ell =(1+B)^{-1} \sum_{\ell} \l(\frac{B}{1+B}\r)^\ell  f_\ell$.

We next expand the above equation in powers of $1/K$, keeping only the leading and sub-leading order terms \cite{vanKampen}. We also use the explicit expressions $\langle \ell\rangle_\ell=B$ and $\langle \ell^2\rangle_\ell = B(2B+1)$ for the first two moments of the geometric distribution.

We arrive at the Fokker--Planck equation
\al{
\partial_t p(x,y,t) = {}& -\partial_x \l\{ \l[ \frac{B}{K} \HIB\l(Ky\r) - \gamma_0 x\r]  p\l(x,y,t\r) \r\}  - \partial_x \l\{ \l[ \frac{B}{K} \HIB\l(Kx\r) - \gamma_0 y\r]  p\l(x,y,t\r) \r\}\nonumber \\
						{}&+ \frac{1}{2K} \partial_x^2 \l\{ \l[{ \frac{B\l(2B+1\r)}{K} \HIB\l(Ky\r) + \gamma_0 x }\r]  p\l(x,y,t\r)  \r\} \nonumber\\
						{}&+ \frac{1}{2K} \partial_y^2 \l\{ \l[{ \frac{B\l(2B+1\r)}{K} \HIB\l(Kx\r) + \gamma_0 y }\r]  p\l(x,y,t\r)  \r\},
}{eq:FPGB}
where we have written $\partial_x=\frac{\partial}{\partial x}$ and similarly for $\partial_y$.
Realisations of the random process described by the Fokker-Planck equation \eqref{eq:FPGB} can be obtained as the solutions of the coupled It\=o stochastic differential equations
\subeq{
dx_t ={}& v(x_t,y_t) dt + \sqrt{D(x_t,y_t)} dW_t^{(x)}, \\
dy_t ={}& v(x_t,y_t) dt + \sqrt{D(y_t,x_t)} dW_t^{(y)}, 
}{eq:diffapprox}
with the defined drift $v$ and diffusion $D$ given by
\subeq{
v(w,z):={}&B \l(r_0+ \frac{r}{1+z^n}\r)-  \gamma_0 w, \\
D(w,z):={}&  \frac{B}{K} \l[{\l(2B+1\r)} \l(r_0+ \frac{r}{1+z^n}\r) +  \frac{1}{B} \gamma_0 w\r].
}{eq:vD}
The quantities $W_t^{(x)}$ and $W_t^{(y)}$ are independent Wiener processes.

\subsection{Diffusion approximation of the CB and NB models}
The same procedure can be applied to the master equation of the CB and NB models, and for completeness we report the resulting Fokker-Planck equations.

For the CB model one finds
\al{
\partial_t p(x,y,t) = {}& -\partial_x \l\{ \l[ \frac{B}{K} \HSIBB \l(Ky\r) - \gamma_0 x\r]  p\l(x,y,t\r) \r\}  - \partial_x \l\{ \l[ \frac{B}{K} \HSIBB\l(Kx\r) - \gamma_0 y\r]  p\l(x,y,t\r) \r\}\nonumber \\
						{}&+ \frac{B}{2K} \partial_{x}^2 \l\{ \l[{ \frac{B}{K} \HSIBB\l(Ky\r) + \frac{1}{B} \gamma_0 x }\r]  p\l(x,y,t\r)  \r\} \nonumber\\
						{}&+ \frac{B}{2K} \partial_{y}^2 \l\{ \l[{ \frac{B}{K} \HSIBB\l(Kx\r) + \frac{1}{B} \gamma_0 y }\r]  p\l(x,y,t\r)  \r\},
}{eq:FPCB}
and for the model without bursts (NB) one has
\al{
\partial_t p(x,y,t) = {}& -\partial_x \l\{ \l[ \frac{1}{K} \HSIB \l(Ky\r) - \gamma_0 x\r]  p\l(x,y,t\r) \r\}  - \partial_x \l\{ \l[ \frac{1}{K} \HSIB\l(Kx\r) - \gamma_0 y\r]  p\l(x,y,t\r) \r\}\nonumber \\
						{}&+ \frac{1}{2K} \partial_{x}^2 \l\{ \l[{ \frac{1}{K} \HSIB\l(Ky\r) +  \gamma_0 x }\r]  p\l(x,y,t\r)  \r\} \nonumber\\
						{}&+ \frac{1}{2K} \partial_{y}^2 \l\{ \l[{ \frac{1}{K} \HSIB\l(Kx\r) + \gamma_0 y }\r]  p\l(x,y,t\r)  \r\}.
}{eq:FPNB}

\section{The piecewise deterministic Markov process (PDMP)}
\subsection{Construction of the PDMP model}
In this section we outline the construction of the PDMP approximation, starting from the full model. For this purpose it is useful to introduce the notation
\eq{
P^{\l(a,b\r)}\l(c,d,t\r) = P(\NMI=a,\NMII=b,\NPI=c,\NPII=d,t).
}{}
Thus the upper indices $(a,b)$ denote the number of mRNA molecules of either type in the system, and the arguments $a, b$ stand for protein numbers.

The master equation \eqref{eq:masterTSIB} can then be written in matrix form
\al{
\frac{d}{dt} \l[
\begin{array}{c}
P^{\l(0,0\r)}\l(c,d,t\r)\\
P^{\l(1,0\r)}\l(c,d,t\r)\\
P^{\l(0,1\r)}\l(c,d,t\r)\\
\ldots
\end{array}
\r] = \mathcal{L}^\dagger \l[
\begin{array}{c}
P^{\l(0,0\r)}\l(c,d,t\r) \\
P^{\l(1,0\r)}\l(c,d,t\r) \\
P^{\l(0,1\r)}\l(c,d,t\r) \\
\ldots
\end{array}
\r]. 
}{eq:Yu}
We have introduced
\eq{
\mathcal{L}^{\dagger}:=\l[
\begin{array}{cccc}
\mathcal{L}^{\dagger\l(0,0\r)} - \HIB(c) - \HIB(d) &  \gamma  & \gamma &   \ldots \\
\HIB(d) &  \mathcal{L}^{\dagger\l(1,0\r)} - \HIB(c) - \HIB(d)-\gamma  & 0   &   \ldots \\
\HIB(c) &  0 & \mathcal{L}^{\dagger\l(0,1\r)} - \HIB(c) - \HIB(d) -\gamma  &   \ldots \\
\ldots &&&
\end{array}
\r] ,
}{}
where the operator $L^{\dagger\l(m,n\r)}$ describes the forward evolution of protein numbers when there are $M_X=m$ and $M_Y=n$ mRNA molecules in the system. Specifically,
\eq{
\mathcal{L}^{\dagger\l(m,n\r)} = m \gamma B\l(\mathcal{E}^{-1,0}-1\r) + n\gamma B\l(\mathcal{E}^{0,-1}-1\r) + \gamma_0 \l[\l(\mathcal{E}^{0,1}-1\r) + \l(\mathcal{E}^{0,1}-1\r)\r],
}{}
where $\mathcal{E}^{i,j}$ are the shift operators \cite{vanKampen} acting on functions of protein numbers. They are defined through 
\eq{
\mathcal{E}^{i,j} f(m,n) \equiv  f(m+i,n+j). 
}{}

Next we consider the limit of fast mRNA decay, that is, large values of $\gamma$. More specifically mRNA molecules of either type are generated with rates $H(N_Y)$ and $H(N_X)$ respectively, and we assume that $\gamma$ is much larger than either of these two rates ($\gamma\gg H(N_Y), \gamma \gg H(N_X)$, for any values of $N_X$ and $N_Y$). In this limit, the system is almost always in the state without mRNA molecules  $(i.e., \NMI=0,\NMII=0)$, except for short spells during which there is either one molecule of mRNA of type $X$, or one of type $Y$.  The duration of the episodes spent in these $(1,0)$ and $(0,1)$ states is of order $\gamma^{-1}$, then a switch back to the $(0,0)$ state occurs. The probability to find the system in states with $M_X>1$ or $M_Y>1$ is even smaller, specifically of order $(H/\gamma)^2$, and we neglect contributions from these states.
Equation \eqref{eq:Yu} can then be simplified into a forward equation of a three-state model:
\al{
\frac{d}{dt} \l[
\begin{array}{c}
P^{\l(0,0\r)}\l(c,d,t\r) \\
P^{\l(1,0\r)}\l(c,d,t\r) \\
P^{\l(0,1\r)}\l(c,d,t\r) \\
\end{array}
\r]  = \mathcal{L^{\dagger}_{\rm{approx}}}
\l[
\begin{array}{c}
P^{\l(0,0\r)} \l(c,d,t\r)\\
P^{\l(1,0\r)} \l(c,d,t\r)\\
P^{\l(0,1\r)} \l(c,d,t\r)\\
\end{array}
\r]
}{eq:Yu2}
with
\eq{
\mathcal{L}^{\dagger}_{\rm{approx}}:=\l[
\begin{array}{ccc}
\mathcal{L}^{\dagger\l(0,0\r)} - \HIB(c) - \HIB(d) &  \gamma  & \gamma  \\
\HIB(d) & \mathcal{L}^{\dagger\l(1,0\r)} -\gamma  & 0    \\
\HIB(c) &  0 & \mathcal{L}^{\dagger\l(0,1\r)}-\gamma  \\
\end{array}
\r].
}{}
In the next step we consider the limit of large values of $K$. Formally we take the limit $K\to\infty$. The system can then be described by the protein concentration $x=\NPI/K$ and $y=\NPII/K$. The corresponding probability distributions in the continuum limit are  
\subeq{
\po(x,y):=P^{(0,0)}\l(Kx, Ky\r) K^2, \\
\pX(x,y):=P^{(1,0)}\l(Kx, Ky\r) K^2, \\
\pY(x,y):=P^{(0,1)}\l(Kx, Ky\r) K^2. 
}{}
On the left-hand side we have introduced the notation $0, X$ and $Y$ to describe the states in which there are no mRNA molecules ($M_X=M_Y=0$), one mRNA molecule of type $X$ ($M_X=1, M_Y=0$) and one mRNA molecule of type $Y$ respectively ($M_X=0, M_Y=1$). This is in-line with the notation in the main manuscript.

The time evolution of the protein concentrations between the random switching events between these three states is then taken to be deterministic.  Mathematically this corresponds to expanding the discrete operators $\mathcal{L}^{\dagger\l(a,b\r)}$ in powers of $K^{-1}$, and keeping only the lowest-order advection terms. This generates so-called Liouville operators, and leads to

\eq{
\frac{\partial}{\partial t} \l[
\begin{array}{c}
\po\\\pX\\\pY
\end{array}
\r]
= \l(L^{\dagger}_{d} + L^{\dagger}_\text{s}\r) \l[
\begin{array}{c}
\po\\\pX\\\pY
\end{array}
\r],
}{eq:Yu3}
where $L^{\dagger}_d$ and $L^{\dagger}_s$ are the forward operators driving the deterministic flow and the random switching between states, respectively. They are given by 
\subeq{
 L^{\dagger}_d :={}& \l[
\begin{array}{ccc}
\l(L^\dagger_d\r)_{11} & 0 & 0\\
0 &\l(L^\dagger_d\r)_{22} & 0 \\
0&0 & \l(L^\dagger_d\r)_{33}
\end{array}
\r], \\
 L^{\dagger}_s :={}& \l[
\begin{array}{ccc}
- \HP(Kx)-\HP(Ky)  & \gamma & \gamma\\
\HP(Ky) &- \gamma & 0 \\
 \HP(Kx)&0 & - \gamma
\end{array}
\r],
}{eq:PDMP operators}
and
\subeq{
\l(L^\dagger_d\r)_{11} :={}&\gamma_0\partial_x \l(x\r) + \gamma_0\partial_y \l(y\r),\\
\l(L^\dagger_d\r)_{22} :={}& \partial_x \l(-\gamma b + \gamma_0 x\r) + \gamma_0 \partial_y \l(y\r) ,\\
\l(L^\dagger_d\r)_{33} :={}&\gamma_0 \partial_x \l(x\r) + \partial_y \l(-\gamma b + \gamma_0y\r).
}{}

\noindent\underline{Nature of the approximation}\\
 In deriving Eq. (\ref{eq:Yu3}) we have made several assumptions and approximations:
 \begin{enumerate}
 \item[(i)] First, we have assumed that $\gamma/H\gg 1$, where $H$ stands for the maximum value $H(\NPI)$ and $H(\NPII)$ can attain. We recall that $H(N)=K\left[r_0+\frac{r}{1+(N/K)^h}\right]$. The function $h(x)=r_0+r/(1+x^n)$ does not involve $K$ or $\gamma$, and its maximum value is $r_0+r$. In dimensionless units, the assumption $\gamma/H\gg 1$ is thus fulfilled if $\gamma\gg (r_0+r)K$.
  \item[(ii)] We have replaced the discrete operators $L^{\l(a,b\r)}$ by deterministic Liouville operators, i.e., we neglected \emph{demographic stochasticity} of the protein degradation. The purpose of this is to isolate the contribution of the bursting noise, originating from the random switching of the mRNA state ($0, X$ and $Y$). Making the deterministic approximation for the protein concentrations is formally valid only in the limit of very large protein populations, $K\gg 1$ ($K$ sets the scale of the numbers of protein molecules).
  \end{enumerate}
In summary we assume $\gamma \gg (r_0+r)K$ and $K\gg 1$. We expect our approximations to be accurate when both of these are fulfilled, in particular the typical value of $\gamma$ above which our theory can be expected to be accurate will depend on the choice of $K$, which in turn must be chosen large enough to justify the deterministic approximation of the protein dynamics. 

The data in the main manuscript reveals that the mathematical approximation agrees well with simulations for $\gamma=30$ and $K=200$. In our simulations we use $r_0\approx 0.007$ and $r=0.06$. 

\subsection{Forward equation in the limit $\gamma \rightarrow \infty$}
We start from
\subeq{
\partial_t \po ={}& \l[\gamma_0 \partial_x  x+ \gamma_0 \partial_y y- \HP(Kx) -\HP(Ky)\r]\po + \gamma \pX + \gamma \pY, \label{eq:alice}\\
\partial_t \pX ={}& \l[\partial_x  \l(\gamma_0 x - \gamma b\r) + \gamma_0 \partial_y - \gamma \r]\pX + \HP\l(Ky\r) \po,\\
\partial_t \pY ={}& \l[\partial_y  \l(\gamma_0 y - \gamma b\r) + \gamma_0 \partial_x - \gamma \r]\pY + \HP\l(Kx\r) \po.
}{}
Applying the operator
$\l[\partial_y  \l(\frac{\gamma_0}{\gamma} y - b\r) + \frac{\gamma_0}{\gamma} \partial_x - 1 \r]\l[\partial_x  \l(\frac{\gamma_0}{\gamma} x - b\r) +\frac{\gamma_0}{\gamma}\partial_y - 1 \r]$ to both sides of equation \eqref{eq:alice} results in
\al{
\partial_t {}&\l[\partial_y  \l(\frac{\gamma_0}{\gamma} y - b\r) + \frac{\gamma_0}{\gamma} \partial_x - 1 \r]\l[\partial_x  \l(\frac{\gamma_0}{\gamma} x - b\r) +\frac{\gamma_0}{\gamma}\partial_y - 1 \r] \po  \nonumber\\
={}& \l[\partial_y  \l(\frac{\gamma_0}{\gamma} y - b\r) + \frac{\gamma_0}{\gamma} \partial_x - 1 \r] \l[\partial_t \pX - \HP\l(Ky\r) \po \r] + \l[\partial_x  \l(\frac{\gamma_0}{\gamma} x - b\r) +\frac{\gamma_0}{\gamma}\partial_y - 1 \r] \l[\partial_t \pY - \HP\l(Kx\r) \po \r]\nonumber\\
{}&+\l[\partial_y  \l(\frac{\gamma_0}{\gamma} y - b\r) + \frac{\gamma_0}{\gamma} \partial_x - 1 \r]\l[\partial_x  \l(\frac{\gamma_0}{\gamma} x - b\r) +\frac{\gamma_0}{\gamma}\partial_y - 1 \r] \l[\gamma_0 \partial_x  x+ \gamma_0 \partial_y y- \HP(Kx) -\HP(Ky)\r]\po.
}{eq:forward finite gamma}
We note that this equation is not closed in $\po$. 

Next, we take the $\gamma \rightarrow \infty$ limit, keeping in mind that $H$ and $\gamma_0$ are finite. The system then almost-surely stays in the 0-state, and consequently $\pX, \pY \rightarrow 0$. Equation \eqref{eq:forward finite gamma} then reduces to
\al{
\partial_t {}&\l(-b\partial_y - 1 \r)\l(-b\partial_x -1 \r) \po =\l(b\partial_y +1 \r) \l[\HP\l(Ky\r) \po \r] +\l(b\partial_x +1 \r) \l[\HP\l(Kx\r) \po \r].
}{eq:forward inf}
The inverse operator of $1+b\partial_z$ is
\eq{
(1+b\partial_z)^{-1} f(z) = \int^z \frac{e^{-\frac{z-z'}{b}}}{b} f(z') dz,
}{}
and so equation \eqref{eq:forward inf} turns into the `forward equation' presented in the main text:
\al{
\partial_t p_0={}& \partial_x \l( \gamma_0 x p_0 \r) + \partial_y \l( \gamma_0 y p_0\r) - \l[ \HP(Kx) + \HP(Ky) \r] p_0 \nonumber\\
{}&+ \HP(Ky) \int_0^{x} \frac{1}{b} e^{-\frac{x-x'}{b}} p_0\l(x',y,t\r)dx'  +  \HP(Kx) \int_0^{y} \frac{1}{b} e^{-\frac{y-y'}{b}} p_0\l(x,y',t\r) dy'.
}{eq:PDMP ID}

\subsection{Equations for the mean first switching time}
Here we illustrate the detail derivation to the adjoint equation in the main text.
We focus on initial conditions $y>x$ and our goal is to calculate the mean time it takes the dynamics to reach states with $x=y$. We write $T_Z(x,y)$ for the time it takes the dynamics to reach a state in which $x=y$ if started from initial condition $x,y$, and in mRNA state $\text{Z} \in\{0,\text{X},\text{Y}\}$. 
The $T_\text{Z}(x,y)$ then satisfy the following adjoint equation \cite{vanKampen}
\eq{
-\l[
\begin{array}{c}
1\\ 1\\ 1
\end{array}
\r]
= \l(L_{d} + L_s\r) \l[
\begin{array}{c}
T_0(x,y) \\T_\text{X}(x,y)\\T_\text{Y}(x,y)
\end{array}
\r],
}{eq:backward eq}
where $L_d$ and $L_s$ are the adjoint operators of $L^\dagger_d$ and $L^\dagger_s$. They are given by
\al{
 L_d :={}& \l[
\begin{array}{ccc}
\l(L_d\r)_{11} & 0 & 0\\
0 &\l(L_d\r)_{22} & 0 \\
0&0 & \l(L_d\r)_{33}
\end{array}
\r]  \text{ and }
 L_s = \l[
\begin{array}{ccc}
- \HP(Kx)-\HP(Ky)  & \HP(Ky) &  \HP(Kx)\\
 \gamma &- \gamma & 0 \\
\gamma &0 & - \gamma
\end{array}
\r],
}{eq:b-operators}
with
\subeq{
\l(L_d\r)_{11} ={}&-\gamma_0 x\partial_x  - \gamma_0 y\partial_y ,\\
\l(L_d\r)_{22} ={}&\l(\gamma b - \gamma_0 x\r) \partial_x  - \gamma_0 \l(y\r) \partial_y,\\
\l(L_d\r)_{33} ={}&-\gamma_0 x \partial_x  + \l(\gamma b - \gamma_0y\r)  \partial_y .
}{}
In the infinitely fast degrading mRNA limit, $\gamma\rightarrow \infty$, equations \eqref{eq:backward eq} can be seen to converge to
\al{
-\l[
\begin{array}{c}
1\\ 0\\ 0
\end{array}
\r]
={}& \l[
\begin{array}{ccc}
\begin{array}{c}
-\gamma_0 x \partial_x  -\gamma_0 y\partial_y 
- \HP(Kx)-\HP(Ky) 
\end{array} & \HP(Ky) & \HP(Kx)\\
1 & b\partial_x  -1  & 0 \\
1&0 & b\partial_y -1 
\end{array}
\r]. \l[
\begin{array}{c}
T_0(x,y) \\T_\text{X}(x,y)\\T_\text{Y}(x,y)
\end{array}
\r], 
}{eq:backward eq2}

The boundary conditions for the mean first exist times are determined by $T_\text{Z}\l(x_b,y_b\r)=0$, for all locations $(x_b, y_b)\in \partial \Omega$ at which the deterministic flow driven by $L_d^\dagger$ flows \emph{out of the domain $\Omega$} in state Z. Next, we specify a bounded domain $\Omega_C := \l\{\l(x,y\r): 0<x<y, y<C \r\}$.
The boundary conditions of equations \eqref{eq:backward eq2} are then $T_{\rm X}(z, z) =0$ and $T_{\rm Y}(z,C)=0 \ \forall z<C$. We now use these boundary conditions, and integrate the second and the third components of the expression in equation \eqref{eq:backward eq2}. Subsequently we send $C\rightarrow \infty$ and arrive at
\al{
-1 ={}& \l[-\gamma_0 x \partial_x  -\gamma_0 y\partial_y - H(Kx)-H(Ky)\r] T_0\l(x,y\r) \nonumber\\
{}&+ H(Ky) \int_x^y \frac{e^{-\frac{x'-x}{b}}}{b}T_0\l(x',y\r) dx'+  H(Kx) \int_y^\infty \frac{e^{-\frac{y'-y}{b}}}{b} T_0\l(x,y'\r) dx'.
}{eq:integro-diff}

\section{WKB analysis}
\subsection{WKB ansatz}
In order to find the quasi-stationary distribution of the PDMP model and of the diffusion approximation of the GB and CB models, one uses the ansatz 
\eq{
p_{\rm stat} = \exp \l\{-\frac{1}{\epsilon} \l[S_0\l(x,y\r) + \mathcal{O}\l(\frac{B}{K}\r)\r]\r\}
}{}
where $\epsilon\propto K^{-1}$ is the magnitude of the intrinsic noise in the protein dynamics. For the purposes of the WKB analysis the noise is assumed to be weak, i.e., $\epsilon \ll 1$.

\subsection{DA of the GB model}
In the context of the diffusion approximation of the GB model we use $\epsilon=B/K$. To leading order ($\mathcal{O}\l(K^0\r)$) one finds a Hamilton--Jacobi equation of the form
\eq{
0 = \frac{1}{2} \l(\nabla S_0\r)^{\mathbf T} \mathbf{D} \l(\nabla S_0\r) + \mathbf{v}^\mathbf{T} \cdot \nabla S_0.
}{eq:GaussianHJ}
The vector $\mathbf v$ denotes the deterministic flow
\eq{
\mathbf{v}\l(x,y\r) := \l[\begin{array}{c} \HMF\l(y\r) - \gamma_0 x\\\HMF\l(x\r) - \gamma_0 y \end{array}\r],
}{eq:vvv} 
and the (scaled) diffusion matrix $\mathbf D$ is given by 
\eq{
\mathbf{D}\l(x,y\r) := \l[\begin{array}{cc} D_{11}\l(x,y\r) & 0 \\ 0 & D_{22}\l(x,y\r) \end{array}\r],
}{}
with entries
\subeq{
D_{11}\l(x,y\r) ={}& \frac{2B+1}{K} H(Ky)+\frac{1}{B}\gamma_0 x,\\
D_{22}\l(x,y\r) ={}&\frac{2B+1}{K}H(Kx)+\frac{1}{B}\gamma_0y.
}{}

\subsection{DA of the CB model}
As before we use $\epsilon=B/K$. A similar leading-order calculation delivers the Hamilton--Jacobi equation, which is again of the form described in equation $\eqref{eq:GaussianHJ}$. The only differences are minor modifications in the diffusion matrix, which now has entries
\subeq{
D_{11}\l(x,y\r) ={}& \HMF(Ky)+ \frac{1}{B}\gamma_0x, \\
D_{22}\l(x,y\r) ={}& \HMF(Kx)+ \frac{1}{B}\gamma_0y.
}{}

\subsection{DA of the NB model}
It is now convenient to use $\epsilon=1/K$. Again one finds a Hamilton-Jacobi equation of the form as above. The diffusion matrix now has entries
\subeq{
D_{11}\l(x,y\r) ={}& \HMF(Ky)+\gamma_0x, \\
D_{22}\l(x,y\r) ={}& \HMF(Kx)+\gamma_0y.
}{}

\subsection{PDMP model}
For the PDMP model, similar calculations deliver the Hamilton--Jacobi equation  
\al{
0 ={}& \l[\gamma_0 x - \HMF\l(Ky\r)\r] \partial_x S_0 + \l[\gamma_0 y - \HMF\l(Kx\r)\r]  \partial_y S_0 \nonumber\\
{}&+ \l[\gamma_0 x+\gamma_0 y -\HMF\l(Kx\r) - \HMF\l(Ky\r)\r] \l(\partial_x S_0\r)\l(\partial_y  S_0\r) \nonumber\\
{}&+\gamma_0 x \l(\partial_x S_0\r)^2 + \gamma_0 y \l(\partial_y S_0\r)^2 +\gamma_0 x \l(\partial_x S_0\r)^2\l(\partial_y  S_0\r) +   \gamma_0 y \l(\partial_x S_0\r)\l(\partial_y  S_0\r)^2.
}{eq:HJPDMP}

\section{Numerical methods}
Sample paths of the individual-based processes (FM, CB, NB, and GB) are generated by the standard kinetic Monte Carlo algorithm \cite{Schwartz,Gillespie} implemented in \text{c++}.  The PDMP process is simulated using the algorithm proposed by Bokes et al \cite{Bokes}. Simulations of the diffusion approximations are performed using the standard Euler--Maruyama algorithm with a constant time step $\delta t = 10^{-4}$. In all cases $10^6$ sample paths are simulated for a sufficiently long time to measure stationary distributions. For the mean first switching times, we sample $105$ initial states on a lattice on the domain $0\le \NPI(0)<\NPII(0)\le 700$. For each initial state, we simulate $10^4$ sample paths, each until they cross the boundary $\NPI=\NPII$ to measure the mean first switching times.  

The geometric minimum action method proposed by Heymann and Vanden--Eijnden \cite{Heymann} is implemented using MATLAB R2010a, and used to find the rate function $S_0$ of the WKB method. For each model, we sample at least $150$ end points and solve for the least-action paths, discretized into 257 equidistant points, connecting one of the fixed points and the end point. The final landscapes are generated by linear interpolation of the rate functions so obtained. 

The finite-difference scheme to solve the adjoint equation was implemented in MATLAB R2010a, discretizing the domain $0\leq x,y\leq C=2000$ into $150\times150$ grid points. The adjoint equation is then transformed to a set of $22500$ linear equations, which is solved using a built-in numerical solver in MATLAB R2010a.

\newcommand{\wid}{0.85}

\newpage
\section{Sample paths of the different models}
\subsection{Full model}
\begin{figure}[h!!!]
\begin{center}
\includegraphics[width=\wid\textwidth]{\figureroot 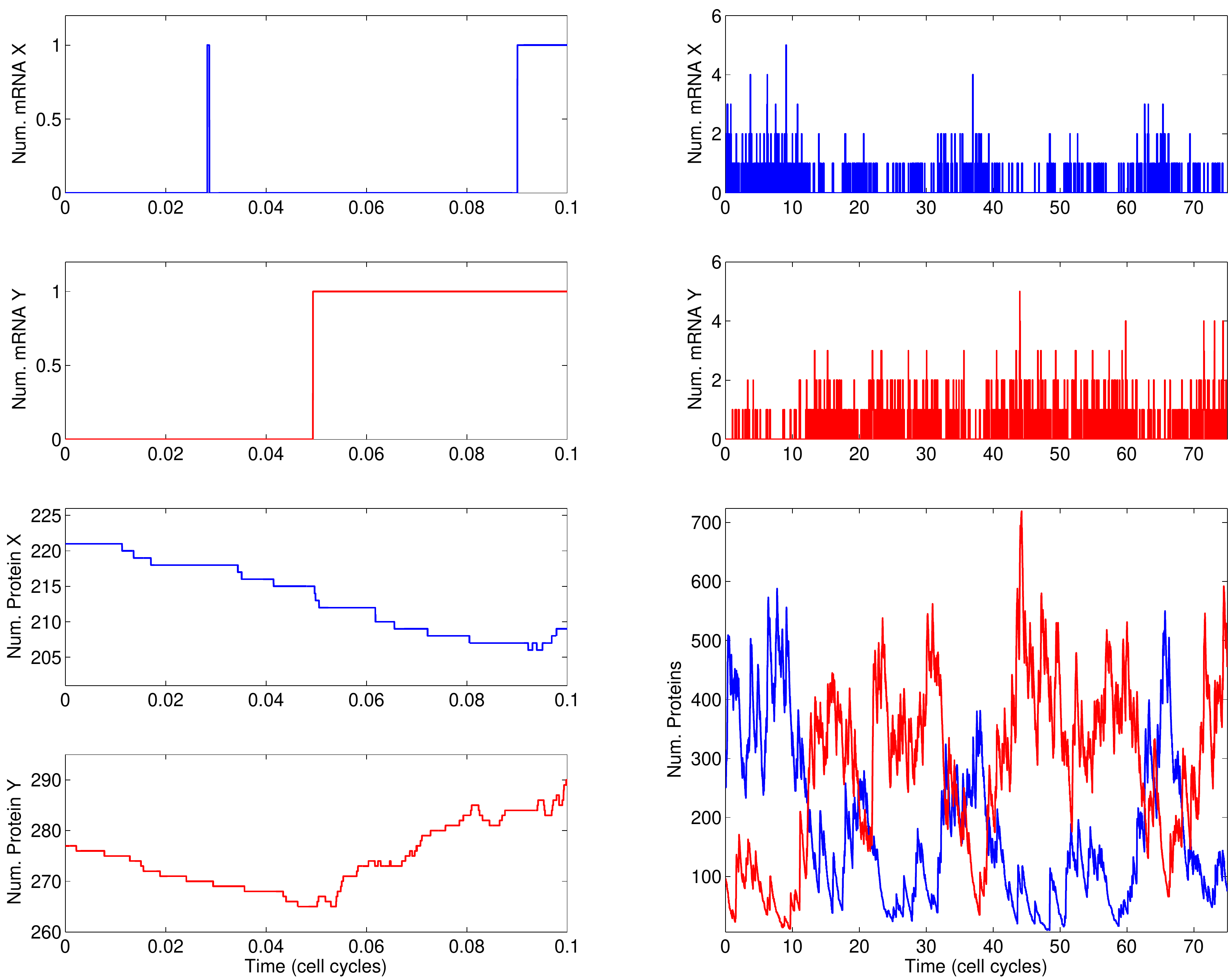}
\caption{One sample path of the full model (FM). Left panel: short time scale. Right panel: the protein expressions switches at a longer time scale driven by intrinsic noise.} \label{fig:S1}
\end{center}
\end{figure}

\newpage
\subsection{Geometrically distributed burst model (GB)}
\begin{figure*}[h!!!]
\begin{center}
\includegraphics[width=\wid\textwidth]{\figureroot 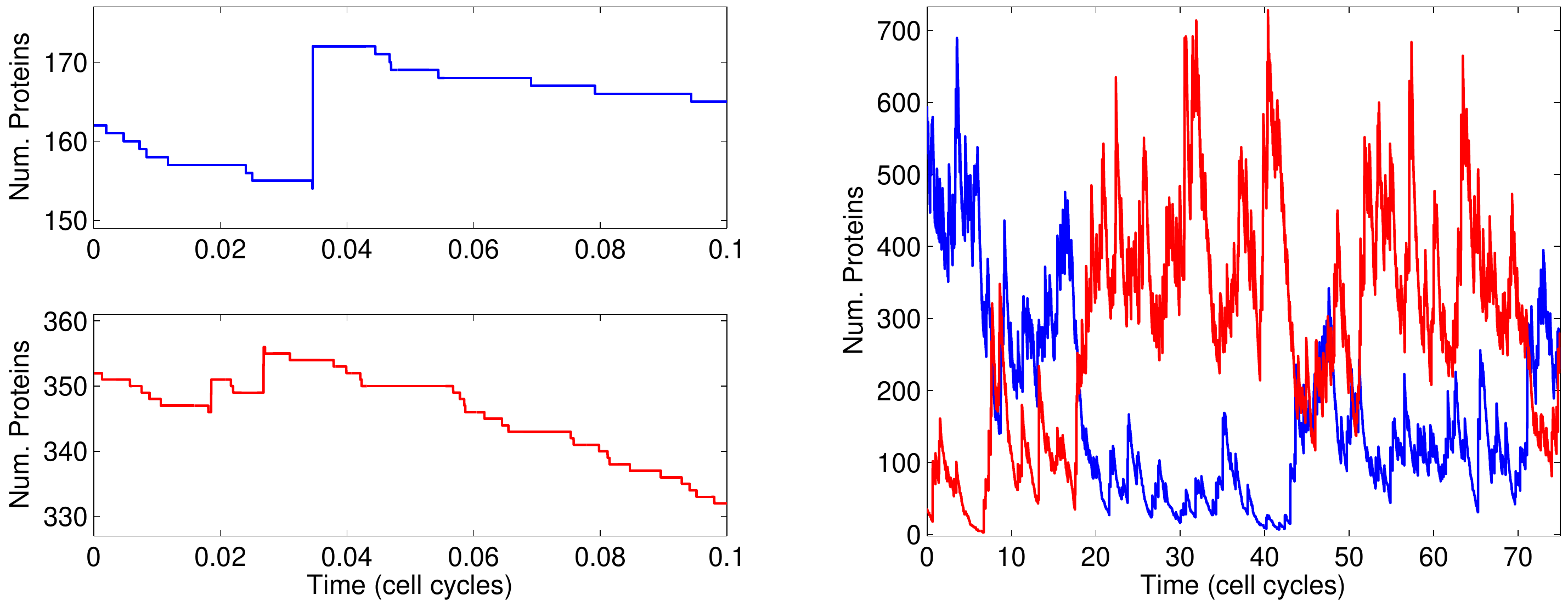}
\caption{One sample path of the model with geometrically distributed burst size (GB). Left panel: short time scale. Right panel: long time scale.} \label{fig:S2}
\end{center}
\end{figure*}

\subsection{Constant burst model (CB)}
\begin{figure*}[h!!!]
\begin{center}
\includegraphics[width=\wid\textwidth]{\figureroot 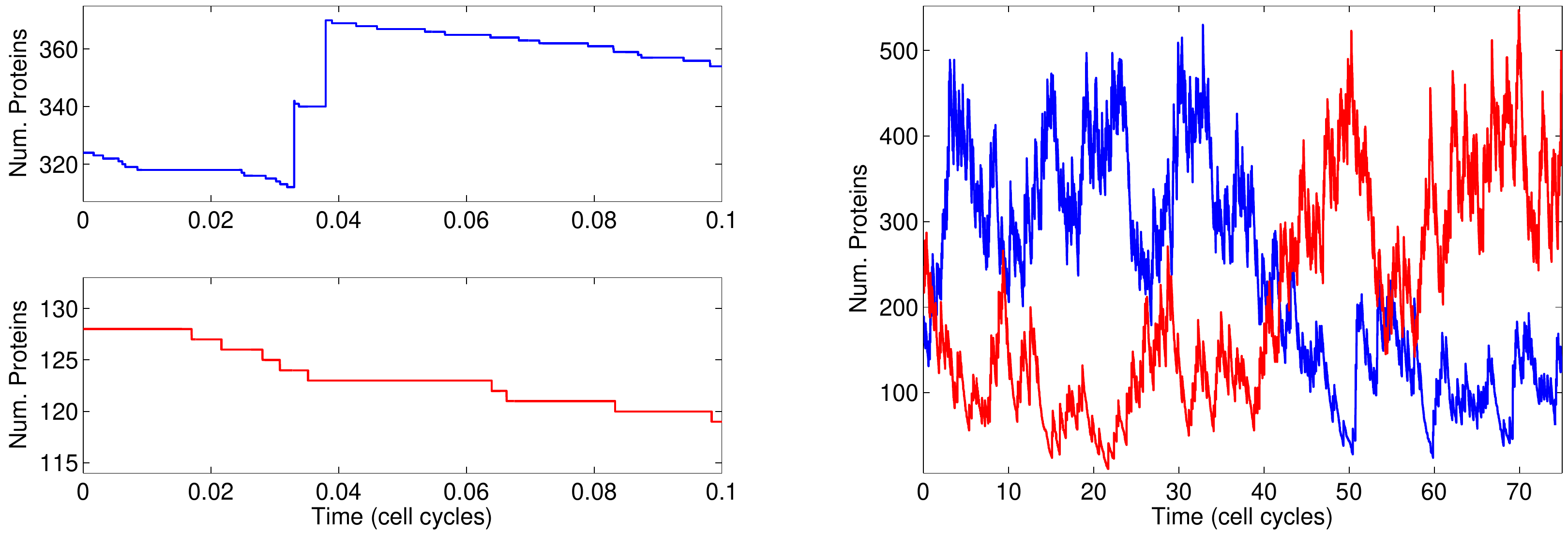}
\caption{One sample path of the model with constant bursts (CB). Left panel: short time scale. Right panel: long time scale.} \label{fig:S2}
\end{center}
\end{figure*}
\newpage

\subsection{No-burst model (NB)}
\begin{figure*}[h!!!]
\begin{center}
\includegraphics[width=\wid\textwidth]{\figureroot 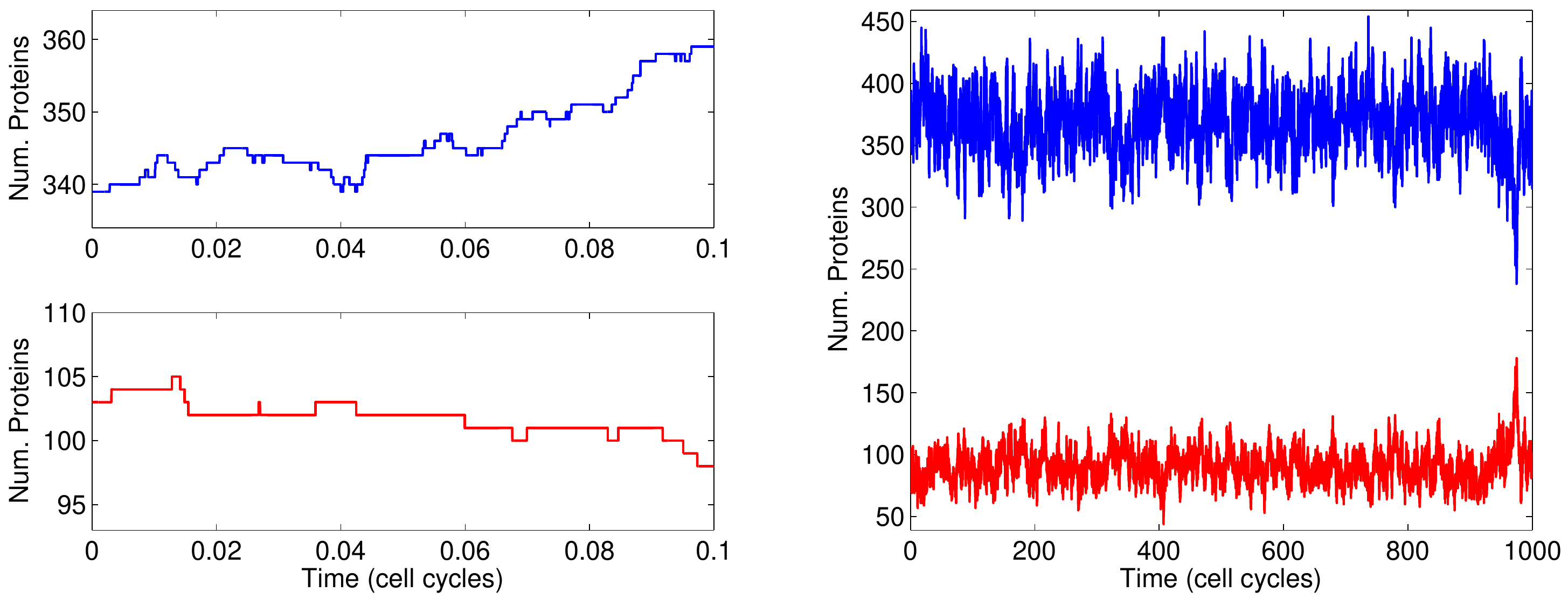}
\caption{One sample path of the model without bursts. Left panel: short time scale. Right panel: long time scale. In $1000$ cell cycles, we observe no switching event in this sample path.} \label{fig:S3}
\end{center}
\end{figure*}

\subsection{PDMP model}
\begin{figure*}[h!!!]
\begin{center}
\includegraphics[width=\wid\textwidth]{\figureroot 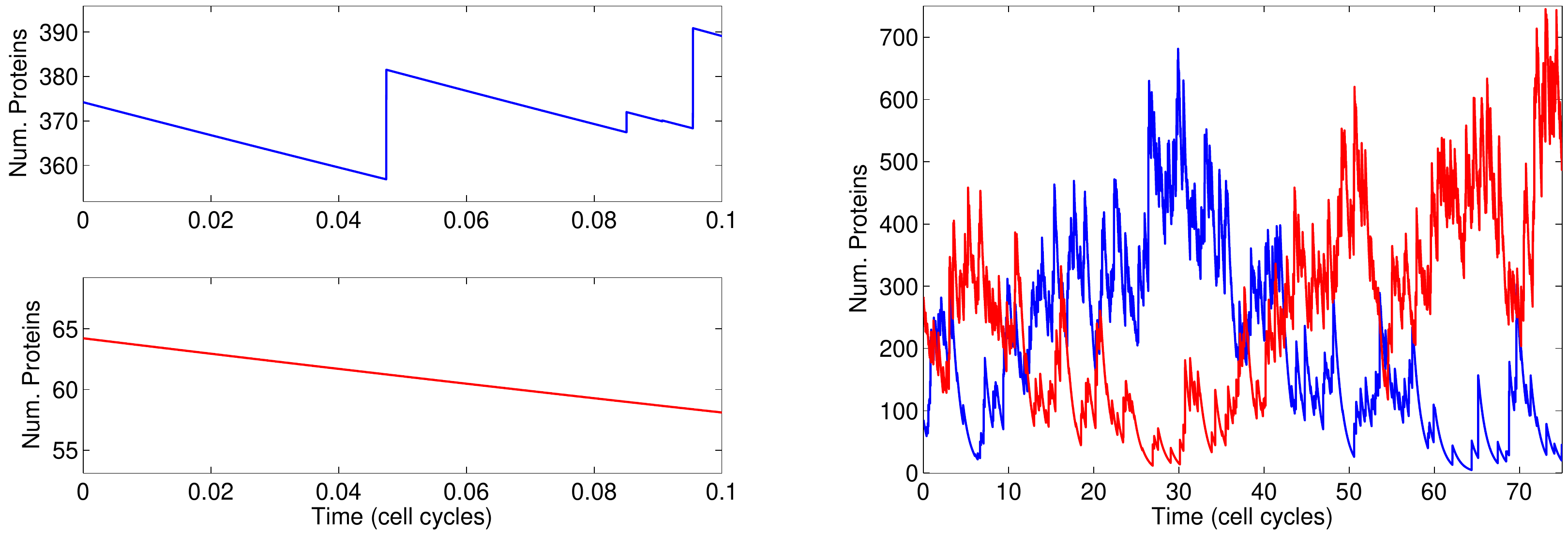}
\caption{One sample path of the PDMP. Left panel: short time scale. Right panel: long time scale.} \label{fig:S4}
\end{center}
\end{figure*}
\newpage

\subsection{Diffusion approximation of the GB model}
\begin{figure*}[h!!!]
\begin{center}
\includegraphics[width=\wid\textwidth]{\figureroot 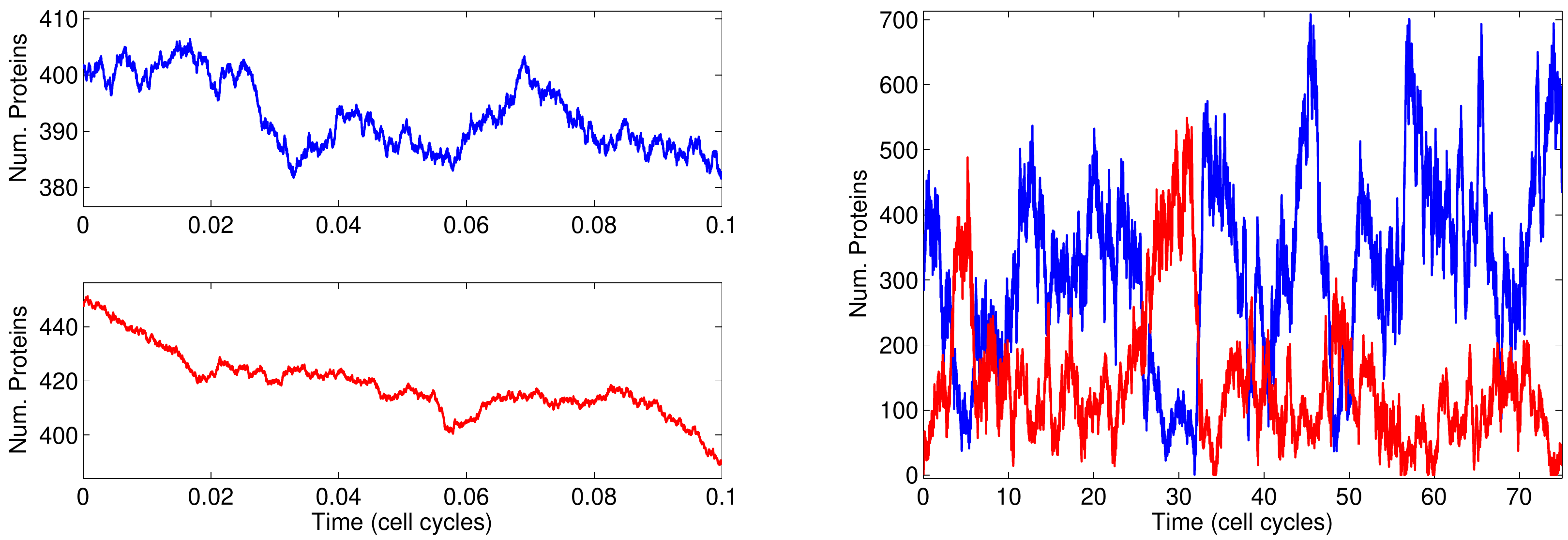}
\caption{One sample path of the diffusion approximation of the GB model. Left panel: short time scale. Right panel: long time scale.} \label{fig:S5}
\end{center}
\end{figure*}

\subsection{Diffusion approximation of the CB model}
\begin{figure*}[h!!!]
\begin{center}
\includegraphics[width=\wid\textwidth]{\figureroot 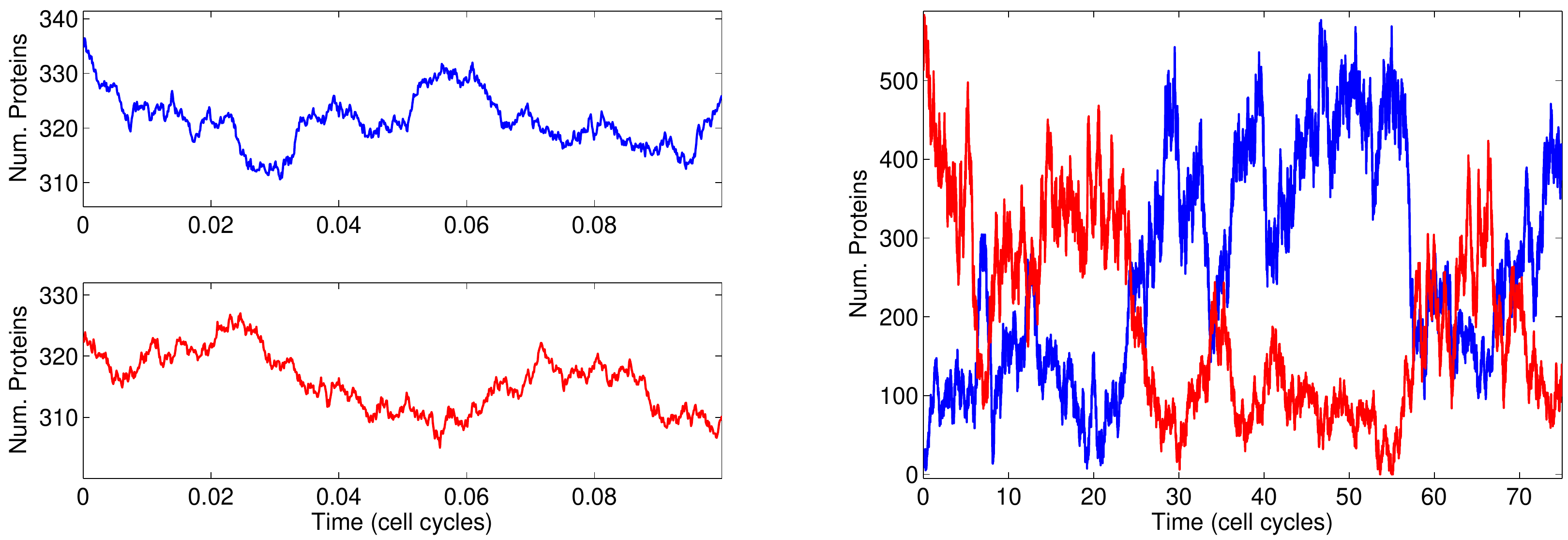}
\caption{One sample path of the diffusion approximation of the CB model. Left panel: short time scale. Right panel: long time scale.} \label{fig:S6}
\end{center}
\end{figure*}

\newpage
\subsection{Diffusion approximation of the NB model}
\begin{figure*}[h!!!]
\begin{center}
\includegraphics[width=\wid\textwidth]{\figureroot 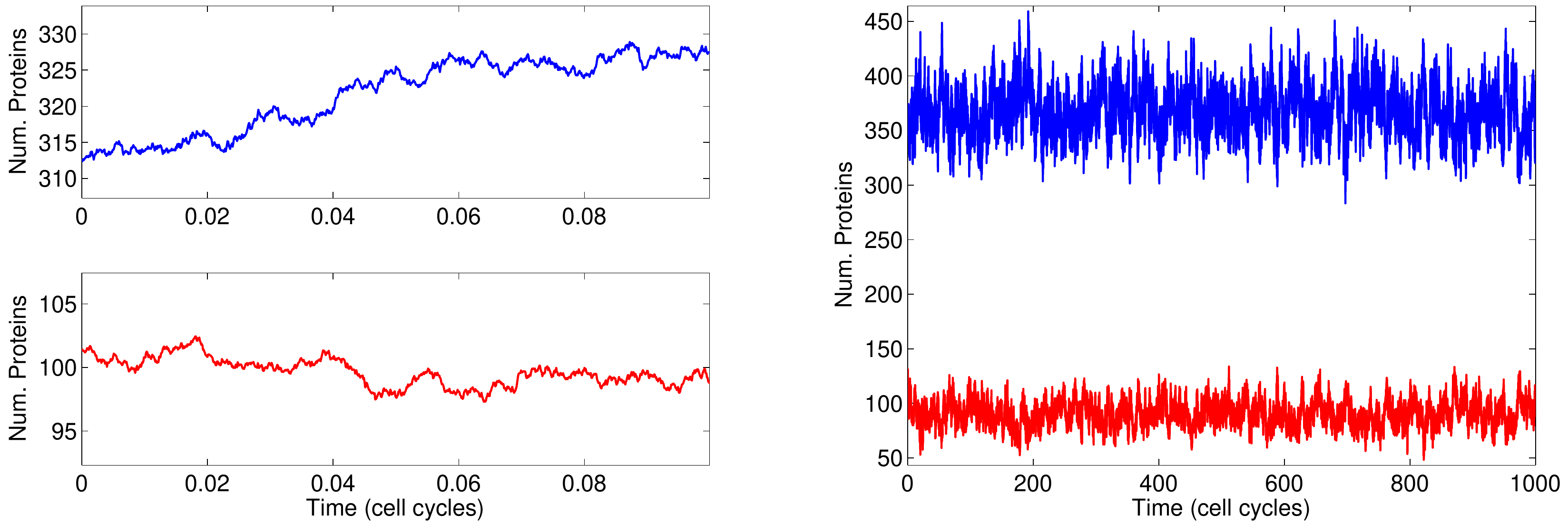}
\caption{One sample path of the diffusion approximation of the single-stage model without bursts. Left panel: short time scale. Right panel: long time scale. Similar to the NB model, no switching event occurs in $1000$ cell cycles in this sample path. } \label{fig:S7}
\end{center}
\end{figure*}

\newpage
\section{Comparison of stationary distributions}
\begin{figure*}[h!!!]
\begin{center}
\includegraphics[width=0.6\textwidth]{\figureroot 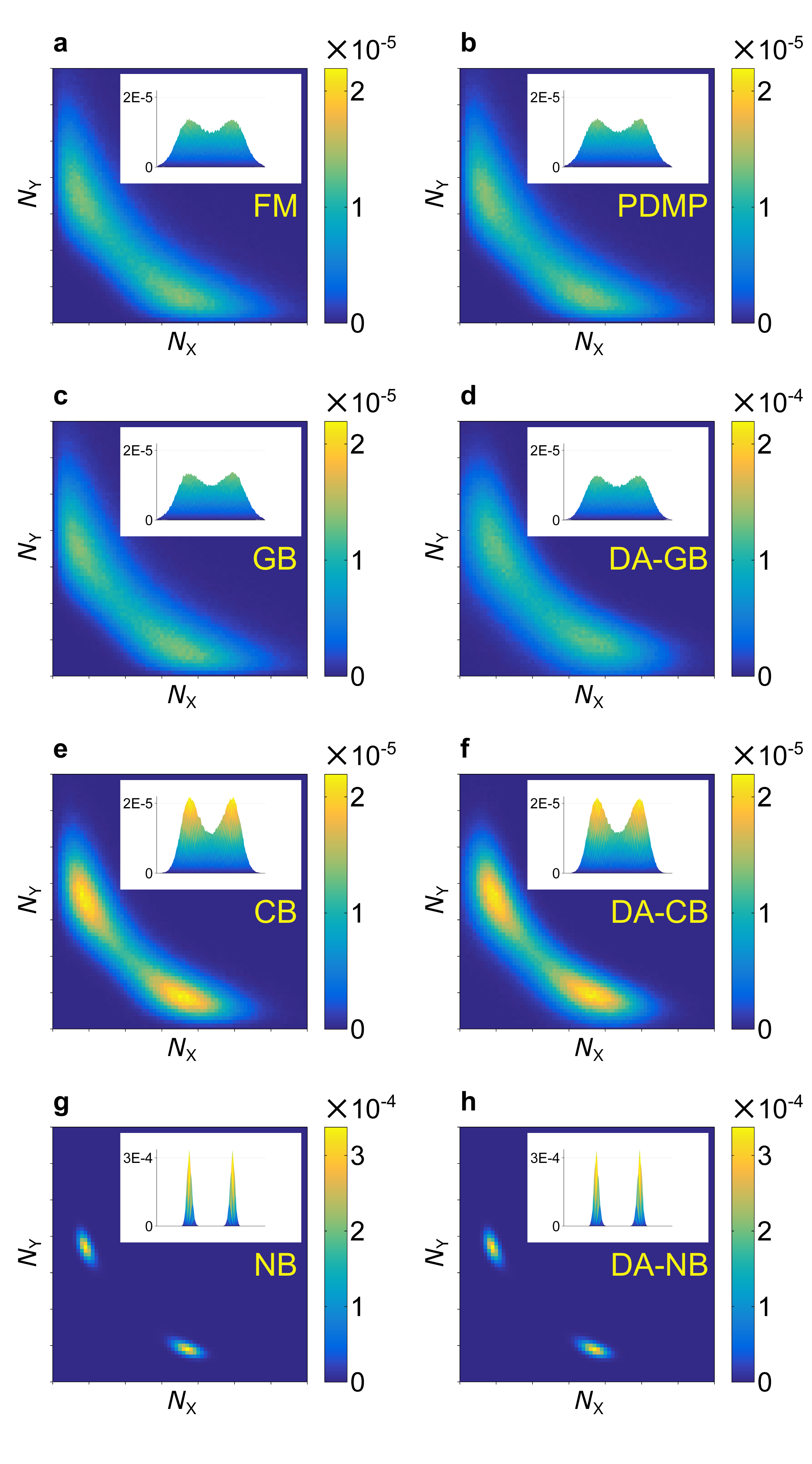}
\caption{Stationary distribution measured in simulations. All axes show $0\leq N_X,N_Y\leq 700$ on a linear scale. Insets show the distribution as viewed from  the point $\NPI=\NPII=700$ facing towards the origin.  ({\bf a}) Full model; ({\bf b}) PDMP; ({\bf c}) GB model; ({\bf d}) Diffusion approximation (DA) of GB; ({\bf e}) CB model; ({\bf f}) DA of CB; ({\bf g}) NB model; ({\bf a}) DA of NB. The same colour scale is used in all panels, except for panels e and f.} 
\end{center}
\end{figure*}

\newpage
\section{Comparison of the mean first switching times}
\begin{figure*}[h!!!]
\begin{center}
\includegraphics[width=0.7\textwidth]{\figureroot 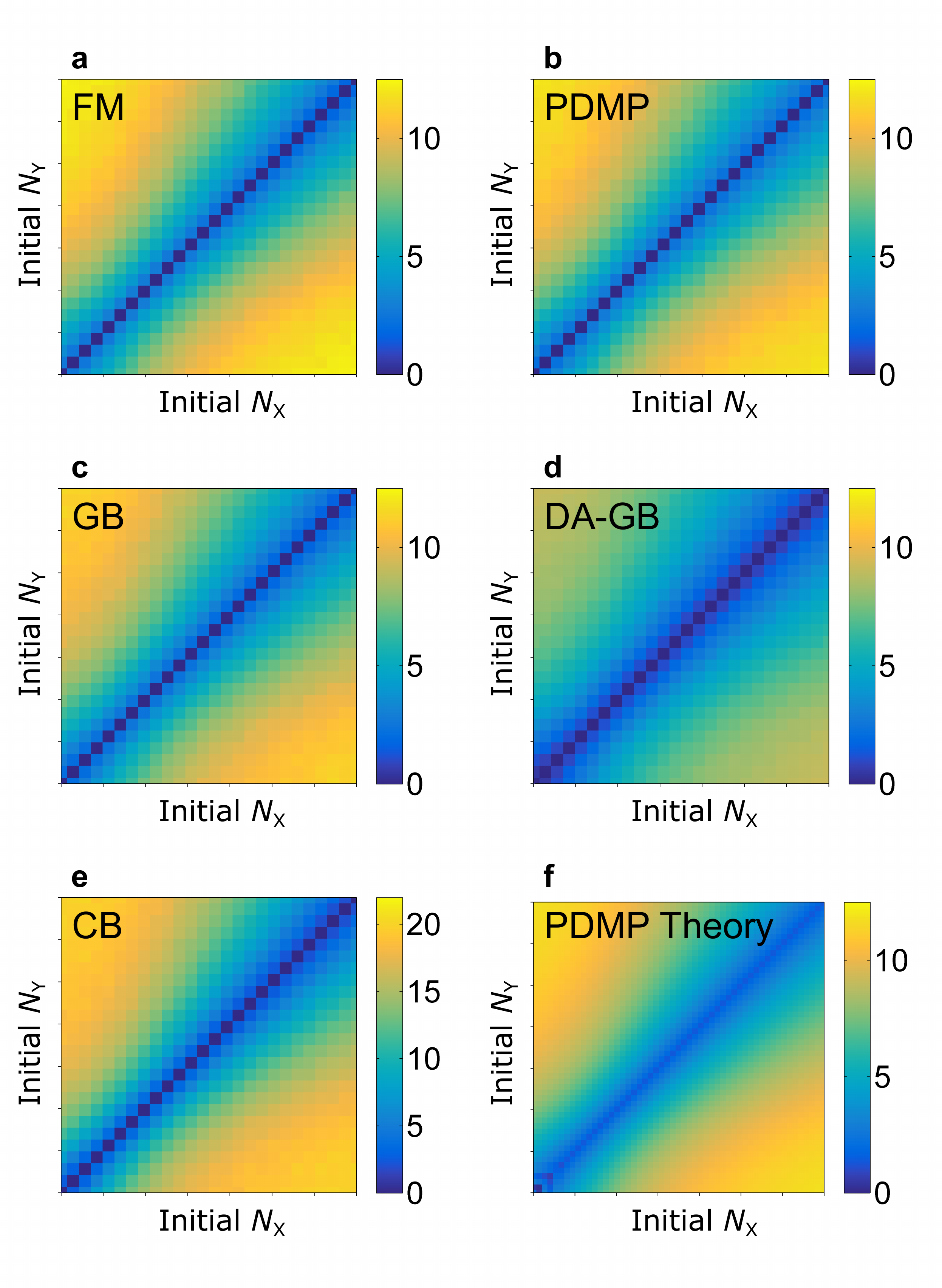}
\caption{Mean first switching times. All graphs show $0\leq N_X,N_Y\leq 700$ on linear scales. ({\bf a}) Full model; ({\bf b}) PDMP; ({\bf c}) GB model; ({\bf d}) Diffusion approximation (DA) of GB; ({\bf e}) CB model; ({\bf f}) Numerical solution of the adjoint equation of the PDMP. Data are plotted on the same colour scale in all panels to allow comparison.} 
\end{center}
\end{figure*}

\newpage
\section{Comparing WKB results}
\begin{figure*}[h!!!]
\begin{center}
\includegraphics[width=0.6\textwidth]{\figureroot 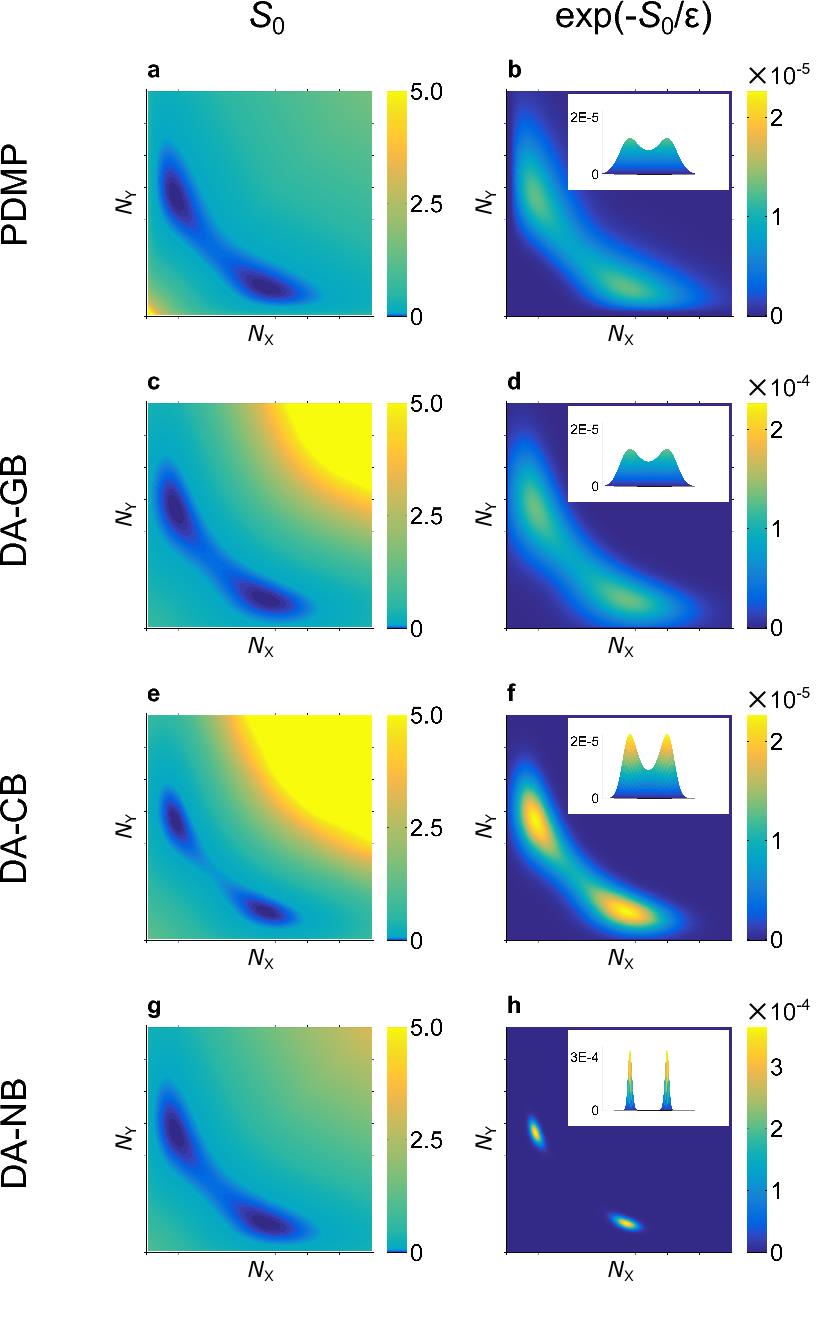}
\caption{Results from the WKB analysis. Panels a, c, e, and g show the WKB rate functions $S_0(N_X,N_Y)$, and panels b, d, f, and h the corresponding approximation for the stationary probability distribution $\mathcal{N} \exp \l[-S_0\l(\NPI,\NPII\r)/\epsilon\r]$ where $\mathcal{N}$ is the normalisation factor. All panels show $0\leq N_X,N_Y\leq 700$ on a linear scale. The insets show the stationary distributions viewed from $\l(\NPI,\NPII\r)=\l(700,700\r)$.  
({\bf a, b}) PDMP; ({\bf c, d}) DA of GB ; ({\bf e, f}) DA of CB; ({\bf g, h}) DA of NB.} 
\end{center}
\end{figure*}

\newpage